\documentclass{article}
\usepackage[utf8]{inputenc}
\usepackage{amsmath}
\usepackage{amssymb}
\usepackage{graphicx}
\usepackage{graphics} % for pdf, bitmapped graphics files
\usepackage{epsfig} % for postscript graphics files

\title{A note on the kinematic model of the planar Purcell's swimmer}
\author{Sudin Kadam, Ravi N. Banavar}
\date{January 2019}

\begin{document}

\maketitle

\begin{figure}[!htb]
\centering
\includegraphics[scale=.36]{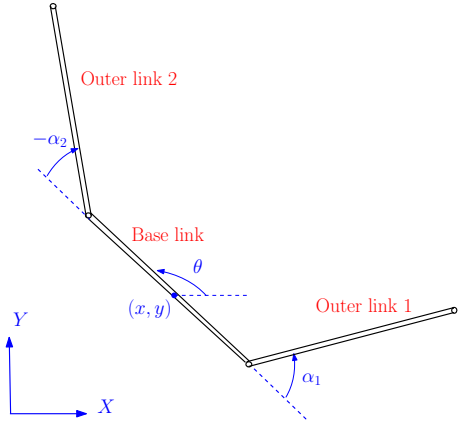}
\caption{The Purcell's swimmer}
\label{Purcell_swimmer}
\end{figure}

We derive the kinematic model of the planar Purcell's swimmer in a stationary fluid based on the approach given in \cite{hatton2013geometric}, \cite{gutman2015symmetries}. The explicit epression for the local form of connection $\mathbb{A}$ is presented. The configuration space of the swimmer is $Q = M \times G$. The swimmer, shown in figure \ref{Purcell_swimmer}, is considered with all 3 links to be slender members with length $2L$. Using Cox theory and the resistive force theory \cite{cox1970motion}, the drag force on an infinitesimal length-element of a slender link is proportional to the local relative fluid velocity. 
We consider body frame attached to each swimmer's center with $x$ axis along the longitudinal direction, $y$ direction along the lateral direction and $\theta$ as the orientation of the x-direction with respect to the reference frame. With $k$ as the viscous drag coefficient, the net forces and moment in the frame body on $i^{th}$ link are 
\begin{align}
    F_{i,x} &= \int_{-L}^{L} \frac{1}{2}k\xi_{i,x}l dl = kL \xi_{i,x} \label{force_x}\\
    F_{i,x} &= \int_{-L}^{L} k\xi_{i,y}l dl = kL \xi_{i,x} \label{force_y} \\
    M_i &= \int_{-L}^{L} k l (l \xi_{i,\theta}l dl = \frac{2}{3}k L^2 \xi_{i, \theta}  \label{moment}
\end{align}
where $F_{i,x}$ and $F_{i,y}$ are respectively the longitudinal and lateral forces, $M_i$ the moment and $\xi_i = [\xi_{i,x}, \xi_{i,y}, \xi_{i,\theta}]^T$ is the body velocity of the $i^{th}$ link.

We refer to the body frame attached to the base link as the base frame. Next, the velocities of the 3 links with respect to the base frame are written in terms of the middle link's velocity and the rate of change of limb angles $\alpha = [\alpha_1, \alpha_2] $ as follows:
\begin{align}
    \xi_1 &= \begin{bmatrix}
cos \alpha_1 \xi_x - sin \alpha_1 \xi_y + sin \alpha_1 L \xi_{\theta} \\
sin \alpha_1 \xi_x + cos \alpha_1 \xi_y  - (1+cos \alpha_1)L \xi_{\theta} + L\dot{\alpha}_1 \\
\xi_{\theta} - \dot{\alpha}_1 \label{zi1}
\end{bmatrix} \\
\xi_2 &= \xi \label{zi2} \\
\xi_3 &= \begin{bmatrix}
cos \alpha_2 \xi_x + sin \alpha_2 \xi_y + sin \alpha_2 L \xi_{\theta} \\
-sin \alpha_2 \xi_x + cos \alpha_2 \xi_y  - (1+cos \alpha_2)L \xi_{\theta} + L\dot{\alpha}_2 \\
\xi_{\theta} + \dot{\alpha}_2
\end{bmatrix} \label{zi3}
\end{align}
The net forces and moment acting on the swimmer can be obtained in the base frame by summing up the forces on all the 3 links using appropriate coordinate transformations as follows
\begin{align}
\begin{bmatrix}
F_x \\
F_y \\
M
\end{bmatrix}  & = \begin{bmatrix}
cos \alpha_1  & sin \alpha_1  & 0 \\
-sin \alpha_1 & cos \alpha_1  & 0\\
L sin \alpha_1  & -L(1+ cos \alpha_1)  & 1
\end{bmatrix}\begin{bmatrix}
F_{1,x} \\
F_{1,y} \\
M_1
\end{bmatrix} + \nonumber \\ 
& \quad \begin{bmatrix}
F_{2,x} \\
F_{2,y} \\
M_2
\end{bmatrix} + \begin{bmatrix}
cos \alpha_2  & -sin \alpha_2  & 0 \\
sin \alpha_2 & cos \alpha_2  & 0\\
L sin \alpha_2  & L(1+ cos \alpha_2)  & 1
\end{bmatrix}\begin{bmatrix}
F_{3,x} \\
F_{3,y} \\
M_3
\end{bmatrix}
\end{align}
Using the expression for individual limb velocities in \eqref{zi1}-\eqref{zi3} and the expressions for forces in \eqref{force_x}-\eqref{moment}, the net forces and moment on the swimmer can be written in terms of $\xi$ and $\dot{\alpha}$ by
\begin{equation}
F = \mathbb{B}^{3 \times 5}(\alpha) \begin{bmatrix}
\xi \\
\dot{\alpha}
\end{bmatrix} = \begin{bmatrix}
\mathbb{B}_1^{3 \times 3}  & \mathbb{B}_2^{3 \times 2}
\end{bmatrix}\begin{bmatrix}
\xi \\
\dot{\alpha}
\end{bmatrix}
\end{equation}

Furthermore, the net drag forces and moments on the entire swimmer go to zero. For $F = [0, 0, 0]^T$, we get $\mathbb{B}_1 \xi = \mathbb{B}_2 \dot{\alpha}$. This leads us to the principal kinematic form of equation in terms of the local connection form $\mathbb{A}(\alpha) = \mathbb{B}_1^{-1} \mathbb{B}_2$, $\mathbb{A}:TM \to \mathfrak{g}$
\begin{equation}
    \xi = \mathbb{A}(\alpha) \dot{\alpha}
\end{equation}

\begin{equation}
\mathbb{A} = \begin{bmatrix} 
             \mathbb{A}_{11}  & \mathbb{A}_{12} \\
             \mathbb{A}_{21}  & \mathbb{A}_{22} \\
             \mathbb{A}_{31}  & \mathbb{A}_{22} 
             \end{bmatrix}
\end{equation}

\begin{align*}
    \mathbb{A}_{11} &= \frac{N_1^{11}}{D_1^{11}} + \frac{N_2^{11}}{D_2^{11}} + \frac{N_3^{11}}{D_3^{11}} \\
    \mathbb{A}_{12} &= \frac{N_1^{12}}{D_1^{12}} + \frac{N_2^{12}}{D_2^{12}} + \frac{N_3^{12}}{D_3^{12}} \\
    \mathbb{A}_{21} &= \frac{N_1^{21}}{D_1^{21}} + \frac{N_2^{21}}{D_2^{21}} + \frac{N_3^{21}}{D_3^{21}} \\
    \mathbb{A}_{22} &= \frac{N_1^{22}}{D_1^{22}} + \frac{N_2^{22}}{D_2^{22}} + \frac{N_3^{22}}{D_3^{22}} \\
    \mathbb{A}_{31} &= \frac{N_1^{31}}{D_1^{31}} + \frac{N_2^{31}}{D_2^{31}} + \frac{N_3^{31}}{D_3^{31}} \\
    \mathbb{A}_{32} &= \frac{N_1^{32}}{D_1^{32}} + \frac{N_2^{32}}{D_2^{32}} + \frac{N_3^{32}}{D_3^{32}} 
\end{align*}
where,
\begin{small}
\begin{align*}
    N_1^{11} &= \left(\frac{2\, L^3\, k}{3} + 2\, L^3\, k\, \left(\cos\!\left(\mathrm{\alpha_1}\right) + 1\right)\right)\, \left(2\, \cos^2 \: \alpha_1\, \cos\!\left(\mathrm{\alpha_2}\right)\, \sin\!\left(\mathrm{\alpha_2}\right) + \cos^2 \: \alpha_1\, \sin\!\left(\mathrm{\alpha_1}\right) + \right.\\
    & \left. \quad 2\, \cos^2 \: \alpha_1\, \sin\!\left(\mathrm{\alpha_2}\right) + 2\, \cos\!\left(\mathrm{\alpha_1}\right)\, \cos^2 \: \alpha_2\, \sin\!\left(\mathrm{\alpha_1}\right) + \cos\!\left(\mathrm{\alpha_1}\right)\, \cos\!\left(\mathrm{\alpha_2}\right)\, \sin\!\left(\mathrm{\alpha_1}\right) + \right.\\
    & \left. \quad  \cos\!\left(\mathrm{\alpha_1}\right)\, \cos\!\left(\mathrm{\alpha_2}\right)\, \sin\!\left(\mathrm{\alpha_2}\right) + \cos\!\left(\mathrm{\alpha_1}\right)\, \sin\!\left(\mathrm{\alpha_1}\right)\, \sin^2 \: \alpha_2 + \right.\\
    & \left. \quad  \cos\!\left(\mathrm{\alpha_1}\right)\, \sin\!\left(\mathrm{\alpha_1}\right) + 2\, \cos^2 \: \alpha_2\, \sin\!\left(\mathrm{\alpha_1}\right) + \cos^2 \: \alpha_2\, \sin\!\left(\mathrm{\alpha_2}\right) +  \cos\!\left(\mathrm{\alpha_2}\right)\, \sin^2 \: \alpha_1\, \sin\!\left(\mathrm{\alpha_2}\right) + \right.\\
    & \left. \quad \cos\!\left(\mathrm{\alpha_2}\right)\, \sin\!\left(\mathrm{\alpha_2}\right) + \sin^3 \: \alpha_1 + \right.\\
    & \left. \quad \sin^2 \: \alpha_1\, \sin\!\left(\mathrm{\alpha_2}\right) + \sin\!\left(\mathrm{\alpha_1}\right)\, \sin^2 \: \alpha_2 + 2\, \sin\!\left(\mathrm{\alpha_1}\right) + {\sin\!\left(\mathrm{\alpha_2}\right)}^3 + 2\, \sin\!\left(\mathrm{\alpha_2}\right)\right)
\end{align*}

\begin{align*}
    D_1^{11} &= 8\, k\, L^2\, \cos^4 \: \alpha_1\, \cos^2 \: \alpha_2 + 8\, k\, L^2\, \cos^4 \: \alpha_1\, \cos\!\left(\mathrm{\alpha_2}\right) + 4\, k\, L^2\, \cos^4 \: \alpha_1\, \sin^2 \: \alpha_2 + \\
    & \quad 6\, k\, L^2\, \cos^4 \: \alpha_1 + 8\, k\, L^2\, \cos^3 \: \alpha_1\, \cos^2 \: \alpha_2 + 4\, k\, L^2\, \cos^3 \: \alpha_1\, \cos\!\left(\mathrm{\alpha_2}\right) + \\
    & \quad 4\, k\, L^2\, \cos^3 \: \alpha_1\, \sin^2 \: \alpha_2 + 4\, k\, L^2\, \cos^3 \: \alpha_1 + 8\, k\, L^2\, \cos^2 \: \alpha_1\, \cos^4 \: \alpha_2 + \\
    & \quad 8\, k\, L^2\, \cos^2 \: \alpha_1\, \cos^3 \: \alpha_2 + 16\, k\, L^2\, \cos^2 \: \alpha_1\, \cos^2 \: \alpha_2\, \sin^2 \: \alpha_1 + \\
    & \quad 16\, k\, L^2\, \cos^2 \: \alpha_1\, \cos^2 \: \alpha_2\, \sin^2 \: \alpha_2 + 20\, k\, L^2\, \cos^2 \: \alpha_1\, \cos^2 \: \alpha_2 + \\
    & \quad 16\, k\, L^2\, \cos^2 \: \alpha_1\, \cos\!\left(\mathrm{\alpha_2}\right)\, \sin^2 \: \alpha_1 - 4\, k\, L^2\, \cos^2 \: \alpha_1\, \cos\!\left(\mathrm{\alpha_2}\right)\, \sin\!\left(\mathrm{\alpha_1}\right)\, \sin\!\left(\mathrm{\alpha_2}\right) + \\
    & \quad 8\, k\, L^2\, \cos^2 \: \alpha_1\, \cos\!\left(\mathrm{\alpha_2}\right)\, \sin^2 \: \alpha_2 + 12\, k\, L^2\, \cos^2 \: \alpha_1\, \cos\!\left(\mathrm{\alpha_2}\right) + \\
    & \quad 8\, k\, L^2\, \cos^2 \: \alpha_1\, \sin^2 \: \alpha_1\, \sin^2 \: \alpha_2 + 12\, k\, L^2\, \cos^2 \: \alpha_1\, \sin^2 \: \alpha_1 - \\
    & \quad 4\, k\, L^2\, \cos^2 \: \alpha_1\, \sin\!\left(\mathrm{\alpha_1}\right)\, \sin\!\left(\mathrm{\alpha_2}\right) + 8\, k\, L^2\, \cos^2 \: \alpha_1\, {\sin\!\left(\mathrm{\alpha_2}\right)}^4 + 16\, k\, L^2\, \cos^2 \: \alpha_1\, \sin^2 \: \alpha_2 + \\
    & \quad 12\, k\, L^2\, \cos^2 \: \alpha_1 + 8\, k\, L^2\, \cos\!\left(\mathrm{\alpha_1}\right)\, \cos^4 \: \alpha_2 + 4\, k\, L^2\, \cos\!\left(\mathrm{\alpha_1}\right)\, \cos^3 \: \alpha_2 + \\
    & \quad 8\, k\, L^2\, \cos\!\left(\mathrm{\alpha_1}\right)\, \cos^2 \: \alpha_2\, \sin^2 \: \alpha_1 - 4\, k\, L^2\, \cos\!\left(\mathrm{\alpha_1}\right)\, \cos^2 \: \alpha_2\, \sin\!\left(\mathrm{\alpha_1}\right)\, \sin\!\left(\mathrm{\alpha_2}\right) + \\
    & \quad 16\, k\, L^2\, \cos\!\left(\mathrm{\alpha_1}\right)\, \cos^2 \: \alpha_2\, \sin^2 \: \alpha_2 + 12\, k\, L^2\, \cos\!\left(\mathrm{\alpha_1}\right)\, \cos^2 \: \alpha_2 + 4\, k\, L^2\, \cos\!\left(\mathrm{\alpha_1}\right)\, \cos\!\left(\mathrm{\alpha_2}\right)\, \sin^2 \: \alpha_1 - \\
    & \quad 4\, k\, L^2\, \cos\!\left(\mathrm{\alpha_1}\right)\, \cos\!\left(\mathrm{\alpha_2}\right)\, \sin\!\left(\mathrm{\alpha_1}\right)\, \sin\!\left(\mathrm{\alpha_2}\right) + 4\, k\, L^2\, \cos\!\left(\mathrm{\alpha_1}\right)\, \cos\!\left(\mathrm{\alpha_2}\right)\, \sin^2 \: \alpha_2 + \\
    & \quad 4\, k\, L^2\, \cos\!\left(\mathrm{\alpha_1}\right)\, \cos\!\left(\mathrm{\alpha_2}\right) + 4\, k\, L^2\, \cos\!\left(\mathrm{\alpha_1}\right)\, \sin^2 \: \alpha_1\, \sin^2 \: \alpha_2 + 4\, k\, L^2\, \cos\!\left(\mathrm{\alpha_1}\right)\, \sin^2 \: \alpha_1 - \\
    & \quad 4\, k\, L^2\, \cos\!\left(\mathrm{\alpha_1}\right)\, \sin\!\left(\mathrm{\alpha_1}\right)\, {\sin\!\left(\mathrm{\alpha_2}\right)}^3 - 4\, k\, L^2\, \cos\!\left(\mathrm{\alpha_1}\right)\, \sin\!\left(\mathrm{\alpha_1}\right)\, \sin\!\left(\mathrm{\alpha_2}\right) + 8\, k\, L^2\, \cos\!\left(\mathrm{\alpha_1}\right)\, {\sin\!\left(\mathrm{\alpha_2}\right)}^4 + \\
    & \quad 12\, k\, L^2\, \cos\!\left(\mathrm{\alpha_1}\right)\, \sin^2 \: \alpha_2 + 4\, k\, L^2\, \cos\!\left(\mathrm{\alpha_1}\right) + 4\, k\, L^2\, \cos^4 \: \alpha_2\, \sin^2 \: \alpha_1 + 6\, k\, L^2\, \cos^4 \: \alpha_2 + \\
    & \quad 4\, k\, L^2\, \cos^3 \: \alpha_2\, \sin^2 \: \alpha_1 + 4\, k\, L^2\, \cos^3 \: \alpha_2 + 8\, k\, L^2\, \cos^2 \: \alpha_2\, \sin^4 \: \alpha_1 + \\
    & \quad 8\, k\, L^2\, \cos^2 \: \alpha_2\, \sin^2 \: \alpha_1\, \sin^2 \: \alpha_2 + 16\, k\, L^2\, \cos^2 \: \alpha_2\, \sin^2 \: \alpha_1 - \\
    & \quad 4\, k\, L^2\, \cos^2 \: \alpha_2\, \sin\!\left(\mathrm{\alpha_1}\right)\, \sin\!\left(\mathrm{\alpha_2}\right) + 12\, k\, L^2\, \cos^2 \: \alpha_2\, \sin^2 \: \alpha_2 + 12\, k\, L^2\, \cos^2 \: \alpha_2 + \\
    & \quad 8\, k\, L^2\, \cos\!\left(\mathrm{\alpha_2}\right)\, \sin^4 \: \alpha_1 - 4\, k\, L^2\, \cos\!\left(\mathrm{\alpha_2}\right)\, \sin^3 \: \alpha_1\, \sin\!\left(\mathrm{\alpha_2}\right) + 4\, k\, L^2\, \cos\!\left(\mathrm{\alpha_2}\right)\, \sin^2 \: \alpha_1\, \sin^2 \: \alpha_2 + \\
    & \quad 12\, k\, L^2\, \cos\!\left(\mathrm{\alpha_2}\right)\, \sin^2 \: \alpha_1 - 4\, k\, L^2\, \cos\!\left(\mathrm{\alpha_2}\right)\, \sin\!\left(\mathrm{\alpha_1}\right)\, \sin\!\left(\mathrm{\alpha_2}\right) + 4\, k\, L^2\, \cos\!\left(\mathrm{\alpha_2}\right)\, \sin^2 \: \alpha_2 + \\
    & \quad  4\, k\, L^2\, \cos\!\left(\mathrm{\alpha_2}\right) + 4\, k\, L^2\, \sin^4 \: \alpha_1\, \sin^2 \: \alpha_2 + 6\, k\, L^2\, \sin^4 \: \alpha_1 - 4\, k\, L^2\, \sin^3 \: \alpha_1\, \sin\!\left(\mathrm{\alpha_2}\right) + \\
    & \quad 4\, k\, L^2\, \sin^2 \: \alpha_1\, {\sin\!\left(\mathrm{\alpha_2}\right)}^4 + 14\, k\, L^2\, \sin^2 \: \alpha_1\, \sin^2 \: \alpha_2 + 12\, k\, L^2\, \sin^2 \: \alpha_1 - 4\, k\, L^2\, \sin\!\left(\mathrm{\alpha_1}\right)\, {\sin\!\left(\mathrm{\alpha_2}\right)}^3 - \\
    & \quad 8\, k\, L^2\, \sin\!\left(\mathrm{\alpha_1}\right)\, \sin\!\left(\mathrm{\alpha_2}\right) + 6\, k\, L^2\, {\sin\!\left(\mathrm{\alpha_2}\right)}^4 + 12\, k\, L^2\, \sin^2 \: \alpha_2 + 6\, k\, L^2
\end{align*}

\begin{align*}
    N_2^{11} &= - 2\, L^2\, k\, \sin\!\left(\mathrm{\alpha_1}\right)\, \left(4\, \cos^2 \: \alpha_1\, \cos^2 \: \alpha_2 + 4\, \cos^2 \: \alpha_1\, \cos\!\left(\mathrm{\alpha_2}\right) + 2\, \cos^2 \: \alpha_1\, \sin^2 \: \alpha_2 + 3\, \cos^2 \: \alpha_1 + \right.\\
    & \left. \quad 4\, \cos\!\left(\mathrm{\alpha_1}\right)\, \cos^2 \: \alpha_2 + 2\, \cos\!\left(\mathrm{\alpha_1}\right)\, \cos\!\left(\mathrm{\alpha_2}\right) + 2\, \cos\!\left(\mathrm{\alpha_1}\right)\, \sin^2 \: \alpha_2 + 2\, \cos\!\left(\mathrm{\alpha_1}\right) + 2\, \cos^2 \: \alpha_2\, \sin^2 \: \alpha_1 + \right.\\
    & \left. \quad 3\, \cos^2 \: \alpha_2 + 2\, \cos\!\left(\mathrm{\alpha_2}\right)\, \sin^2 \: \alpha_1 + 2\, \cos\!\left(\mathrm{\alpha_2}\right) + \sin^2 \: \alpha_1\, \sin^2 \: \alpha_2 + 2\, \sin^2 \: \alpha_1 + 2\, \sin^2 \: \alpha_2 + 3\right)
\end{align*}

\begin{align*}
    D_2^{11} &= 4\, L\, k\, \cos^4 \: \alpha_1\, \cos^2 \: \alpha_2 + 4\, L\, k\, \cos^4 \: \alpha_1\, \cos\!\left(\mathrm{\alpha_2}\right) + 2\, L\, k\, \cos^4 \: \alpha_1\, \sin^2 \: \alpha_2 + 3\, L\, k\, \cos^4 \: \alpha_1 + \\
    & \quad 4\, L\, k\, \cos^3 \: \alpha_1\, \cos^2 \: \alpha_2 + 2\, L\, k\, \cos^3 \: \alpha_1\, \cos\!\left(\mathrm{\alpha_2}\right) + 2\, L\, k\, \cos^3 \: \alpha_1\, \sin^2 \: \alpha_2 + 2\, L\, k\, \cos^3 \: \alpha_1 + \\
    & \quad 4\, L\, k\, \cos^2 \: \alpha_1\, \cos^4 \: \alpha_2 + 4\, L\, k\, \cos^2 \: \alpha_1\, \cos^3 \: \alpha_2 + 8\, L\, k\, \cos^2 \: \alpha_1\, \cos^2 \: \alpha_2\, \sin^2 \: \alpha_1 + \\
    & \quad 8\, L\, k\, \cos^2 \: \alpha_1\, \cos^2 \: \alpha_2\, \sin^2 \: \alpha_2 + 10\, L\, k\, \cos^2 \: \alpha_1\, \cos^2 \: \alpha_2 + 8\, L\, k\, \cos^2 \: \alpha_1\, \cos\!\left(\mathrm{\alpha_2}\right)\, \sin^2 \: \alpha_1 - \\
    & \quad 2\, L\, k\, \cos^2 \: \alpha_1\, \cos\!\left(\mathrm{\alpha_2}\right)\, \sin\!\left(\mathrm{\alpha_1}\right)\, \sin\!\left(\mathrm{\alpha_2}\right) + 4\, L\, k\, \cos^2 \: \alpha_1\, \cos\!\left(\mathrm{\alpha_2}\right)\, \sin^2 \: \alpha_2 + 6\, L\, k\, \cos^2 \: \alpha_1\, \cos\!\left(\mathrm{\alpha_2}\right) + \\
    & \quad 4\, L\, k\, \cos^2 \: \alpha_1\, \sin^2 \: \alpha_1\, \sin^2 \: \alpha_2 + 6\, L\, k\, \cos^2 \: \alpha_1\, \sin^2 \: \alpha_1 - 2\, L\, k\, \cos^2 \: \alpha_1\, \sin\!\left(\mathrm{\alpha_1}\right)\, \sin\!\left(\mathrm{\alpha_2}\right) + \\
    & \quad 4\, L\, k\, \cos^2 \: \alpha_1\, {\sin\!\left(\mathrm{\alpha_2}\right)}^4 + 8\, L\, k\, \cos^2 \: \alpha_1\, \sin^2 \: \alpha_2 + 6\, L\, k\, \cos^2 \: \alpha_1 + 4\, L\, k\, \cos\!\left(\mathrm{\alpha_1}\right)\, \cos^4 \: \alpha_2 + \\
    & \quad 2\, L\, k\, \cos\!\left(\mathrm{\alpha_1}\right)\, \cos^3 \: \alpha_2 + 4\, L\, k\, \cos\!\left(\mathrm{\alpha_1}\right)\, \cos^2 \: \alpha_2\, \sin^2 \: \alpha_1 - 2\, L\, k\, \cos\!\left(\mathrm{\alpha_1}\right)\, \cos^2 \: \alpha_2\, \sin\!\left(\mathrm{\alpha_1}\right)\, \sin\!\left(\mathrm{\alpha_2}\right) + \\
    & \quad 8\, L\, k\, \cos\!\left(\mathrm{\alpha_1}\right)\, \cos^2 \: \alpha_2\, \sin^2 \: \alpha_2 + 6\, L\, k\, \cos\!\left(\mathrm{\alpha_1}\right)\, \cos^2 \: \alpha_2 + 2\, L\, k\, \cos\!\left(\mathrm{\alpha_1}\right)\, \cos\!\left(\mathrm{\alpha_2}\right)\, \sin^2 \: \alpha_1 - \\
    & \quad 2\, L\, k\, \cos\!\left(\mathrm{\alpha_1}\right)\, \cos\!\left(\mathrm{\alpha_2}\right)\, \sin\!\left(\mathrm{\alpha_1}\right)\, \sin\!\left(\mathrm{\alpha_2}\right) + 2\, L\, k\, \cos\!\left(\mathrm{\alpha_1}\right)\, \cos\!\left(\mathrm{\alpha_2}\right)\, \sin^2 \: \alpha_2 + 2\, L\, k\, \cos\!\left(\mathrm{\alpha_1}\right)\, \cos\!\left(\mathrm{\alpha_2}\right) + \\
    & \quad 2\, L\, k\, \cos\!\left(\mathrm{\alpha_1}\right)\, \sin^2 \: \alpha_1\, \sin^2 \: \alpha_2 + 2\, L\, k\, \cos\!\left(\mathrm{\alpha_1}\right)\, \sin^2 \: \alpha_1 - 2\, L\, k\, \cos\!\left(\mathrm{\alpha_1}\right)\, \sin\!\left(\mathrm{\alpha_1}\right)\, {\sin\!\left(\mathrm{\alpha_2}\right)}^3 - \\
    & \quad 2\, L\, k\, \cos\!\left(\mathrm{\alpha_1}\right)\, \sin\!\left(\mathrm{\alpha_1}\right)\, \sin\!\left(\mathrm{\alpha_2}\right) + 4\, L\, k\, \cos\!\left(\mathrm{\alpha_1}\right)\, {\sin\!\left(\mathrm{\alpha_2}\right)}^4 + 6\, L\, k\, \cos\!\left(\mathrm{\alpha_1}\right)\, \sin^2 \: \alpha_2 + 2\, L\, k\, \cos\!\left(\mathrm{\alpha_1}\right) + \\
    & \quad 2\, L\, k\, \cos^4 \: \alpha_2\, \sin^2 \: \alpha_1 + 3\, L\, k\, \cos^4 \: \alpha_2 + 2\, L\, k\, \cos^3 \: \alpha_2\, \sin^2 \: \alpha_1 + 2\, L\, k\, \cos^3 \: \alpha_2 + \\
    & \quad 4\, L\, k\, \cos^2 \: \alpha_2\, \sin^4 \: \alpha_1 + 4\, L\, k\, \cos^2 \: \alpha_2\, \sin^2 \: \alpha_1\, \sin^2 \: \alpha_2 + 8\, L\, k\, \cos^2 \: \alpha_2\, \sin^2 \: \alpha_1 - \\
    & \quad 2\, L\, k\, \cos^2 \: \alpha_2\, \sin\!\left(\mathrm{\alpha_1}\right)\, \sin\!\left(\mathrm{\alpha_2}\right) + 6\, L\, k\, \cos^2 \: \alpha_2\, \sin^2 \: \alpha_2 + 6\, L\, k\, \cos^2 \: \alpha_2 + 4\, L\, k\, \cos\!\left(\mathrm{\alpha_2}\right)\, \sin^4 \: \alpha_1 - \\
    & \quad 2\, L\, k\, \cos\!\left(\mathrm{\alpha_2}\right)\, \sin^3 \: \alpha_1\, \sin\!\left(\mathrm{\alpha_2}\right) + 2\, L\, k\, \cos\!\left(\mathrm{\alpha_2}\right)\, \sin^2 \: \alpha_1\, \sin^2 \: \alpha_2 + 6\, L\, k\, \cos\!\left(\mathrm{\alpha_2}\right)\, \sin^2 \: \alpha_1 - \\
    & \quad 2\, L\, k\, \cos\!\left(\mathrm{\alpha_2}\right)\, \sin\!\left(\mathrm{\alpha_1}\right)\, \sin\!\left(\mathrm{\alpha_2}\right) + 2\, L\, k\, \cos\!\left(\mathrm{\alpha_2}\right)\, \sin^2 \: \alpha_2 + 2\, L\, k\, \cos\!\left(\mathrm{\alpha_2}\right) + 2\, L\, k\, \sin^4 \: \alpha_1\, \sin^2 \: \alpha_2 + \\
    & \quad 3\, L\, k\, \sin^4 \: \alpha_1 - 2\, L\, k\, \sin^3 \: \alpha_1\, \sin\!\left(\mathrm{\alpha_2}\right) + 2\, L\, k\, \sin^2 \: \alpha_1\, {\sin\!\left(\mathrm{\alpha_2}\right)}^4 + 7\, L\, k\, \sin^2 \: \alpha_1\, \sin^2 \: \alpha_2 + \\
    & \quad 6\, L\, k\, \sin^2 \: \alpha_1 - 2\, L\, k\, \sin\!\left(\mathrm{\alpha_1}\right)\, {\sin\!\left(\mathrm{\alpha_2}\right)}^3 - 4\, L\, k\, \sin\!\left(\mathrm{\alpha_1}\right)\, \sin\!\left(\mathrm{\alpha_2}\right) + \\
    & \quad 3\, L\, k\, {\sin\!\left(\mathrm{\alpha_2}\right)}^4 + 6\, L\, k\, \sin^2 \: \alpha_2 + 3\, L\, k
\end{align*}

\begin{align*}
    N_3^{11} &= - L^2\, k\, \cos\!\left(\mathrm{\alpha_1}\right)\, \left(2\, \cos^2 \: \alpha_1\, \cos\!\left(\mathrm{\alpha_2}\right)\, \sin\!\left(\mathrm{\alpha_2}\right) + \cos^2 \: \alpha_1\, \sin\!\left(\mathrm{\alpha_1}\right) + 2\, \cos^2 \: \alpha_1\, \sin\!\left(\mathrm{\alpha_2}\right) - \right.\\
    & \left. \quad 2\, \cos\!\left(\mathrm{\alpha_1}\right)\, \cos^2 \: \alpha_2\, \sin\!\left(\mathrm{\alpha_1}\right) - 3\, \cos\!\left(\mathrm{\alpha_1}\right)\, \cos\!\left(\mathrm{\alpha_2}\right)\, \sin\!\left(\mathrm{\alpha_1}\right) + 3\, \cos\!\left(\mathrm{\alpha_1}\right)\, \cos\!\left(\mathrm{\alpha_2}\right)\, \sin\!\left(\mathrm{\alpha_2}\right) - \right. \\
    & \left. \quad \cos\!\left(\mathrm{\alpha_1}\right)\, \sin\!\left(\mathrm{\alpha_1}\right)\, \sin^2 \: \alpha_2 - \cos\!\left(\mathrm{\alpha_1}\right)\, \sin\!\left(\mathrm{\alpha_1}\right) + 2\, \cos\!\left(\mathrm{\alpha_1}\right)\, \sin\!\left(\mathrm{\alpha_2}\right) - 2\, \cos^2 \: \alpha_2\, \sin\!\left(\mathrm{\alpha_1}\right) - \right.\\
    & \left. \quad \cos^2 \: \alpha_2\, \sin\!\left(\mathrm{\alpha_2}\right) + \cos\!\left(\mathrm{\alpha_2}\right)\, \sin^2 \: \alpha_1\, \sin\!\left(\mathrm{\alpha_2}\right) - 2\, \cos\!\left(\mathrm{\alpha_2}\right)\, \sin\!\left(\mathrm{\alpha_1}\right) + \cos\!\left(\mathrm{\alpha_2}\right)\, \sin\!\left(\mathrm{\alpha_2}\right) + \right.\\
    & \left. \quad  \sin^3 \: \alpha_1 + \sin^2 \: \alpha_1\, \sin\!\left(\mathrm{\alpha_2}\right) - \sin\!\left(\mathrm{\alpha_1}\right)\, \sin^2 \: \alpha_2 - {\sin\!\left(\mathrm{\alpha_2}\right)}^3\right) \\
D_3^{11} &= D_2^{11}
\end{align*}

\begin{align*}
    N_1^{12} &= \left(\frac{2\, L^3\, k}{3} + 2\, L^3\, k\, \left(\cos\!\left(\mathrm{\alpha_1}\right) + 1\right)\right)\, \left(2\, \cos^4 \: \alpha_1 + 2\, \cos^3 \: \alpha_1 - 2\, \cos^2 \: \alpha_1\, \cos\!\left(\mathrm{\alpha_2}\right) +  \right.\\
    & \left. \quad 4\, \cos^2 \: \alpha_1\, \sin^2 \: \alpha_1 + 3\, \cos^2 \: \alpha_1\, \sin^2 \: \alpha_2 + 2\, \cos^2 \: \alpha_1 + 2\, \cos\!\left(\mathrm{\alpha_1}\right)\, \cos^2 \: \alpha_2 +  \right.\\
    & \left. \quad 2\, \cos\!\left(\mathrm{\alpha_1}\right)\, \sin^2 \: \alpha_1 - 2\, \cos\!\left(\mathrm{\alpha_1}\right)\, \sin\!\left(\mathrm{\alpha_1}\right)\, \sin\!\left(\mathrm{\alpha_2}\right) + 4\, \cos\!\left(\mathrm{\alpha_1}\right)\, \sin^2 \: \alpha_2 + 2\, \cos\!\left(\mathrm{\alpha_1}\right) -  \right.\\
    & \left. \quad 2\, \cos^4 \: \alpha_2 - 2\, \cos^3 \: \alpha_2 - 3\, \cos^2 \: \alpha_2\, \sin^2 \: \alpha_1 - 4\, \cos^2 \: \alpha_2\, \sin^2 \: \alpha_2 - 2\, \cos^2 \: \alpha_2 -  \right.\\
    & \left. \quad 4\, \cos\!\left(\mathrm{\alpha_2}\right)\, \sin^2 \: \alpha_1 + 2\, \cos\!\left(\mathrm{\alpha_2}\right)\, \sin\!\left(\mathrm{\alpha_1}\right)\, \sin\!\left(\mathrm{\alpha_2}\right) - 2\, \cos\!\left(\mathrm{\alpha_2}\right)\, \sin^2 \: \alpha_2 -  \right.\\
    & \left. \quad 2\, \cos\!\left(\mathrm{\alpha_2}\right) + 2\, \sin^4 \: \alpha_1 + \sin^2 \: \alpha_1 - 2\, {\sin\!\left(\mathrm{\alpha_2}\right)}^4 - \sin^2 \: \alpha_2\right)
\end{align*}

\begin{align*}
    D_1^{12} &= 16\, k\, L^2\, \cos^4 \: \alpha_1\, \cos^2 \: \alpha_2 + 16\, k\, L^2\, \cos^4 \: \alpha_1\, \cos\!\left(\mathrm{\alpha_2}\right) +  8\, k\, L^2\, \cos^4 \: \alpha_1\, \sin^2 \: \alpha_2 + \\
    & \quad 12\, k\, L^2\, \cos^4 \: \alpha_1 + 16\, k\, L^2\, \cos^3 \: \alpha_1\, \cos^2 \: \alpha_2 + 8\, k\, L^2\, \cos^3 \: \alpha_1\, \cos\!\left(\mathrm{\alpha_2}\right) + \\
    & \quad 8\, k\, L^2\, \cos^3 \: \alpha_1\, \sin^2 \: \alpha_2 + 8\, k\, L^2\, \cos^3 \: \alpha_1 + 16\, k\, L^2\, \cos^2 \: \alpha_1\, \cos^4 \: \alpha_2 + \\
    & \quad 16\, k\, L^2\, \cos^2 \: \alpha_1\, \cos^3 \: \alpha_2 + 32\, k\, L^2\, \cos^2 \: \alpha_1\, \cos^2 \: \alpha_2\, \sin^2 \: \alpha_1 + \\
    & \quad 32\, k\, L^2\, \cos^2 \: \alpha_1\, \cos^2 \: \alpha_2\, \sin^2 \: \alpha_2 + 40\, k\, L^2\, \cos^2 \: \alpha_1\, \cos^2 \: \alpha_2 + 32\, k\, L^2\, \cos^2 \: \alpha_1\, \cos\!\left(\mathrm{\alpha_2}\right)\, \sin^2 \: \alpha_1 - \\
    & \quad 8\, k\, L^2\, \cos^2 \: \alpha_1\, \cos\!\left(\mathrm{\alpha_2}\right)\, \sin\!\left(\mathrm{\alpha_1}\right)\, \sin\!\left(\mathrm{\alpha_2}\right) + 16\, k\, L^2\, \cos^2 \: \alpha_1\, \cos\!\left(\mathrm{\alpha_2}\right)\, \sin^2 \: \alpha_2 + \\
    & \quad 24\, k\, L^2\, \cos^2 \: \alpha_1\, \cos\!\left(\mathrm{\alpha_2}\right) + 16\, k\, L^2\, \cos^2 \: \alpha_1\, \sin^2 \: \alpha_1\, \sin^2 \: \alpha_2 + 24\, k\, L^2\, \cos^2 \: \alpha_1\, \sin^2 \: \alpha_1 - \\
    & \quad 8\, k\, L^2\, \cos^2 \: \alpha_1\, \sin\!\left(\mathrm{\alpha_1}\right)\, \sin\!\left(\mathrm{\alpha_2}\right) + 16\, k\, L^2\, \cos^2 \: \alpha_1\, {\sin\!\left(\mathrm{\alpha_2}\right)}^4 + 32\, k\, L^2\, \cos^2 \: \alpha_1\, \sin^2 \: \alpha_2 + \\
    & \quad 24\, k\, L^2\, \cos^2 \: \alpha_1 + 16\, k\, L^2\, \cos\!\left(\mathrm{\alpha_1}\right)\, \cos^4 \: \alpha_2 + 8\, k\, L^2\, \cos\!\left(\mathrm{\alpha_1}\right)\, \cos^3 \: \alpha_2 + \\
    & \quad 16\, k\, L^2\, \cos\!\left(\mathrm{\alpha_1}\right)\, \cos^2 \: \alpha_2\, \sin^2 \: \alpha_1 - 8\, k\, L^2\, \cos\!\left(\mathrm{\alpha_1}\right)\, \cos^2 \: \alpha_2\, \sin\!\left(\mathrm{\alpha_1}\right)\, \sin\!\left(\mathrm{\alpha_2}\right) + \\
    & \quad 32\, k\, L^2\, \cos\!\left(\mathrm{\alpha_1}\right)\, \cos^2 \: \alpha_2\, \sin^2 \: \alpha_2 + 24\, k\, L^2\, \cos\!\left(\mathrm{\alpha_1}\right)\, \cos^2 \: \alpha_2 + 8\, k\, L^2\, \cos\!\left(\mathrm{\alpha_1}\right)\, \cos\!\left(\mathrm{\alpha_2}\right)\, \sin^2 \: \alpha_1 - \\
    & \quad 8\, k\, L^2\, \cos\!\left(\mathrm{\alpha_1}\right)\, \cos\!\left(\mathrm{\alpha_2}\right)\, \sin\!\left(\mathrm{\alpha_1}\right)\, \sin\!\left(\mathrm{\alpha_2}\right) + 8\, k\, L^2\, \cos\!\left(\mathrm{\alpha_1}\right)\, \cos\!\left(\mathrm{\alpha_2}\right)\, \sin^2 \: \alpha_2 + \\
    & \quad 8\, k\, L^2\, \cos\!\left(\mathrm{\alpha_1}\right)\, \cos\!\left(\mathrm{\alpha_2}\right) + 8\, k\, L^2\, \cos\!\left(\mathrm{\alpha_1}\right)\, \sin^2 \: \alpha_1\, \sin^2 \: \alpha_2 + \\
    & \quad 8\, k\, L^2\, \cos\!\left(\mathrm{\alpha_1}\right)\, \sin^2 \: \alpha_1 - 8\, k\, L^2\, \cos\!\left(\mathrm{\alpha_1}\right)\, \sin\!\left(\mathrm{\alpha_1}\right)\, {\sin\!\left(\mathrm{\alpha_2}\right)}^3 - 8\, k\, L^2\, \cos\!\left(\mathrm{\alpha_1}\right)\, \sin\!\left(\mathrm{\alpha_1}\right)\, \sin\!\left(\mathrm{\alpha_2}\right) + \\
    & \quad 16\, k\, L^2\, \cos\!\left(\mathrm{\alpha_1}\right)\, {\sin\!\left(\mathrm{\alpha_2}\right)}^4 + 24\, k\, L^2\, \cos\!\left(\mathrm{\alpha_1}\right)\, \sin^2 \: \alpha_2 + 8\, k\, L^2\, \cos\!\left(\mathrm{\alpha_1}\right) + 8\, k\, L^2\, \cos^4 \: \alpha_2\, \sin^2 \: \alpha_1 + \\
    & \quad 12\, k\, L^2\, \cos^4 \: \alpha_2 + 8\, k\, L^2\, \cos^3 \: \alpha_2\, \sin^2 \: \alpha_1 + 8\, k\, L^2\, \cos^3 \: \alpha_2 + \\
    & \quad 16\, k\, L^2\, \cos^2 \: \alpha_2\, \sin^4 \: \alpha_1 + 16\, k\, L^2\, \cos^2 \: \alpha_2\, \sin^2 \: \alpha_1\, \sin^2 \: \alpha_2 + 32\, k\, L^2\, \cos^2 \: \alpha_2\, \sin^2 \: \alpha_1 - \\
    & \quad 8\, k\, L^2\, \cos^2 \: \alpha_2\, \sin\!\left(\mathrm{\alpha_1}\right)\, \sin\!\left(\mathrm{\alpha_2}\right) + 24\, k\, L^2\, \cos^2 \: \alpha_2\, \sin^2 \: \alpha_2 + 24\, k\, L^2\, \cos^2 \: \alpha_2 + \\
    & \quad 16\, k\, L^2\, \cos\!\left(\mathrm{\alpha_2}\right)\, \sin^4 \: \alpha_1 - 8\, k\, L^2\, \cos\!\left(\mathrm{\alpha_2}\right)\, \sin^3 \: \alpha_1\, \sin\!\left(\mathrm{\alpha_2}\right) + 8\, k\, L^2\, \cos\!\left(\mathrm{\alpha_2}\right)\, \sin^2 \: \alpha_1\, \sin^2 \: \alpha_2 + \\
    & \quad 24\, k\, L^2\, \cos\!\left(\mathrm{\alpha_2}\right)\, \sin^2 \: \alpha_1 - 8\, k\, L^2\, \cos\!\left(\mathrm{\alpha_2}\right)\, \sin\!\left(\mathrm{\alpha_1}\right)\, \sin\!\left(\mathrm{\alpha_2}\right) + 8\, k\, L^2\, \cos\!\left(\mathrm{\alpha_2}\right)\, \sin^2 \: \alpha_2 + \\
    & \quad 8\, k\, L^2\, \cos\!\left(\mathrm{\alpha_2}\right) + 8\, k\, L^2\, \sin^4 \: \alpha_1\, \sin^2 \: \alpha_2 + 12\, k\, L^2\, \sin^4 \: \alpha_1 - 8\, k\, L^2\, \sin^3 \: \alpha_1\, \sin\!\left(\mathrm{\alpha_2}\right) + \\
    & \quad 8\, k\, L^2\, \sin^2 \: \alpha_1\, {\sin\!\left(\mathrm{\alpha_2}\right)}^4 + 28\, k\, L^2\, \sin^2 \: \alpha_1\, \sin^2 \: \alpha_2 + 24\, k\, L^2\, \sin^2 \: \alpha_1 - 8\, k\, L^2\, \sin\!\left(\mathrm{\alpha_1}\right)\, {\sin\!\left(\mathrm{\alpha_2}\right)}^3 - \\
    & \quad 16\, k\, L^2\, \sin\!\left(\mathrm{\alpha_1}\right)\, \sin\!\left(\mathrm{\alpha_2}\right) + 12\, k\, L^2\, {\sin\!\left(\mathrm{\alpha_2}\right)}^4 + 24\, k\, L^2\, \sin^2 \: \alpha_2 + 12\, k\, L^2
\end{align*}
\begin{align*}
    N_2^{12} &= - L^2\, k\, \sin\!\left(\mathrm{\alpha_1}\right)\, \left(2\, \cos^2 \: \alpha_1\, \cos\!\left(\mathrm{\alpha_2}\right)\, \sin\!\left(\mathrm{\alpha_2}\right) +  \cos^2 \: \alpha_1\, \sin\!\left(\mathrm{\alpha_1}\right) + 2\, \cos^2 \: \alpha_1\, \sin\!\left(\mathrm{\alpha_2}\right) -  \right.\\
    & \left. \quad 2\, \cos\!\left(\mathrm{\alpha_1}\right)\, \cos^2 \: \alpha_2\, \sin\!\left(\mathrm{\alpha_1}\right) - 3\, \cos\!\left(\mathrm{\alpha_1}\right)\, \cos\!\left(\mathrm{\alpha_2}\right)\, \sin\!\left(\mathrm{\alpha_1}\right) + 3\, \cos\!\left(\mathrm{\alpha_1}\right)\, \cos\!\left(\mathrm{\alpha_2}\right)\, \sin\!\left(\mathrm{\alpha_2}\right) - \right.\\
    & \left. \quad  \cos\!\left(\mathrm{\alpha_1}\right)\, \sin\!\left(\mathrm{\alpha_1}\right)\, \sin^2 \: \alpha_2 - \cos\!\left(\mathrm{\alpha_1}\right)\, \sin\!\left(\mathrm{\alpha_1}\right) + 2\, \cos\!\left(\mathrm{\alpha_1}\right)\, \sin\!\left(\mathrm{\alpha_2}\right) - 2\, \cos^2 \: \alpha_2\, \sin\!\left(\mathrm{\alpha_1}\right) - \right.\\
    & \left. \quad  \cos^2 \: \alpha_2\, \sin\!\left(\mathrm{\alpha_2}\right) + \cos\!\left(\mathrm{\alpha_2}\right)\, \sin^2 \: \alpha_1\, \sin\!\left(\mathrm{\alpha_2}\right) - 2\, \cos\!\left(\mathrm{\alpha_2}\right)\, \sin\!\left(\mathrm{\alpha_1}\right) + \cos\!\left(\mathrm{\alpha_2}\right)\, \sin\!\left(\mathrm{\alpha_2}\right) + \right.\\
    & \left. \quad  \sin^3 \: \alpha_1 + \sin^2 \: \alpha_1\, \sin\!\left(\mathrm{\alpha_2}\right) - \sin\!\left(\mathrm{\alpha_1}\right)\, \sin^2 \: \alpha_2 - {\sin\!\left(\mathrm{\alpha_2}\right)}^3\right) \\
    D_2^{12} &= D_2^{12} 
\end{align*}
\begin{align*}
    N_3^{12} &= - L^2\, k\, \cos\!\left(\mathrm{\alpha_1}\right)\, \left(2\, \cos^4 \: \alpha_1 + 4\, \cos^3 \: \alpha_1 + 4\, \cos^2 \: \alpha_1\, \cos^2 \: \alpha_2 + 4\, \cos^2 \: \alpha_1\, \cos\!\left(\mathrm{\alpha_2}\right) +  \right.\\
    & \left. \quad 4\, \cos^2 \: \alpha_1\, \sin^2 \: \alpha_1 + 5\, \cos^2 \: \alpha_1\, \sin^2 \: \alpha_2 + 8\, \cos^2 \: \alpha_1 + 4\, \cos\!\left(\mathrm{\alpha_1}\right)\, \cos^2 \: \alpha_2 -  \right.\\
    & \left. \quad 2\, \cos\!\left(\mathrm{\alpha_1}\right)\, \cos\!\left(\mathrm{\alpha_2}\right)\, \sin\!\left(\mathrm{\alpha_1}\right)\, \sin\!\left(\mathrm{\alpha_2}\right) + 4\, \cos\!\left(\mathrm{\alpha_1}\right)\, \sin^2 \: \alpha_1 - 4\, \cos\!\left(\mathrm{\alpha_1}\right)\, \sin\!\left(\mathrm{\alpha_1}\right)\, \sin\!\left(\mathrm{\alpha_2}\right) +  \right.\\
    & \left. \quad 8\, \cos\!\left(\mathrm{\alpha_1}\right)\, \sin^2 \: \alpha_2 + 4\, \cos\!\left(\mathrm{\alpha_1}\right) + 2\, \cos^4 \: \alpha_2 + 4\, \cos^3 \: \alpha_2 + 5\, \cos^2 \: \alpha_2\, \sin^2 \: \alpha_1 +  \right.\\
    & \left. \quad 4\, \cos^2 \: \alpha_2\, \sin^2 \: \alpha_2 + 8\, \cos^2 \: \alpha_2 + 8\, \cos\!\left(\mathrm{\alpha_2}\right)\, \sin^2 \: \alpha_1 - 4\, \cos\!\left(\mathrm{\alpha_2}\right)\, \sin\!\left(\mathrm{\alpha_1}\right)\, \sin\!\left(\mathrm{\alpha_2}\right) +  \right.\\
    & \left. \quad 4\, \cos\!\left(\mathrm{\alpha_2}\right)\, \sin^2 \: \alpha_2 + 4\, \cos\!\left(\mathrm{\alpha_2}\right) + 2\, \sin^4 \: \alpha_1 + 4\, \sin^2 \: \alpha_1\, \sin^2 \: \alpha_2 + 9\, \sin^2 \: \alpha_1 -  \right.\\
    & \left. \quad 8\, \sin\!\left(\mathrm{\alpha_1}\right)\, \sin\!\left(\mathrm{\alpha_2}\right) + 2\, {\sin\!\left(\mathrm{\alpha_2}\right)}^4 + 9\, \sin^2 \: \alpha_2 + 6\right)
\end{align*}
\begin{align*}
    D_3^{12} &= 8\, L\, k\, \cos^4 \: \alpha_1\, \cos^2 \: \alpha_2 + 8\, L\, k\, \cos^4 \: \alpha_1\, \cos\!\left(\mathrm{\alpha_2}\right) + 4\, L\, k\, \cos^4 \: \alpha_1\, \sin^2 \: \alpha_2 + 6\, L\, k\, \cos^4 \: \alpha_1 +  \\
    & \quad 8\, L\, k\, \cos^3 \: \alpha_1\, \cos^2 \: \alpha_2 + 4\, L\, k\, \cos^3 \: \alpha_1\, \cos\!\left(\mathrm{\alpha_2}\right) + 4\, L\, k\, \cos^3 \: \alpha_1\, \sin^2 \: \alpha_2 + 4\, L\, k\, \cos^3 \: \alpha_1 + \\
    & \quad  8\, L\, k\, \cos^2 \: \alpha_1\, \cos^4 \: \alpha_2 + 8\, L\, k\, \cos^2 \: \alpha_1\, \cos^3 \: \alpha_2 + 16\, L\, k\, \cos^2 \: \alpha_1\, \cos^2 \: \alpha_2\, \sin^2 \: \alpha_1 +  \\
    & \quad 16\, L\, k\, \cos^2 \: \alpha_1\, \cos^2 \: \alpha_2\, \sin^2 \: \alpha_2 + 20\, L\, k\, \cos^2 \: \alpha_1\, \cos^2 \: \alpha_2 + 16\, L\, k\, \cos^2 \: \alpha_1\, \cos\!\left(\mathrm{\alpha_2}\right)\, \sin^2 \: \alpha_1 -  \\
    & \quad 4\, L\, k\, \cos^2 \: \alpha_1\, \cos\!\left(\mathrm{\alpha_2}\right)\, \sin\!\left(\mathrm{\alpha_1}\right)\, \sin\!\left(\mathrm{\alpha_2}\right) + 8\, L\, k\, \cos^2 \: \alpha_1\, \cos\!\left(\mathrm{\alpha_2}\right)\, \sin^2 \: \alpha_2 +  \\
    & \quad 12\, L\, k\, \cos^2 \: \alpha_1\, \cos\!\left(\mathrm{\alpha_2}\right) + 8\, L\, k\, \cos^2 \: \alpha_1\, \sin^2 \: \alpha_1\, \sin^2 \: \alpha_2 + 12\, L\, k\, \cos^2 \: \alpha_1\, \sin^2 \: \alpha_1 -  \\
    & \quad 4\, L\, k\, \cos^2 \: \alpha_1\, \sin\!\left(\mathrm{\alpha_1}\right)\, \sin\!\left(\mathrm{\alpha_2}\right) + 8\, L\, k\, \cos^2 \: \alpha_1\, {\sin\!\left(\mathrm{\alpha_2}\right)}^4 + 16\, L\, k\, \cos^2 \: \alpha_1\, \sin^2 \: \alpha_2 +  \\
    & \quad 12\, L\, k\, \cos^2 \: \alpha_1 + 8\, L\, k\, \cos\!\left(\mathrm{\alpha_1}\right)\, \cos^4 \: \alpha_2 + 4\, L\, k\, \cos\!\left(\mathrm{\alpha_1}\right)\, \cos^3 \: \alpha_2 +  \\
    & \quad 8\, L\, k\, \cos\!\left(\mathrm{\alpha_1}\right)\, \cos^2 \: \alpha_2\, \sin^2 \: \alpha_1 - 4\, L\, k\, \cos\!\left(\mathrm{\alpha_1}\right)\, \cos^2 \: \alpha_2\, \sin\!\left(\mathrm{\alpha_1}\right)\, \sin\!\left(\mathrm{\alpha_2}\right) +  \\
    & \quad 16\, L\, k\, \cos\!\left(\mathrm{\alpha_1}\right)\, \cos^2 \: \alpha_2\, \sin^2 \: \alpha_2 + 12\, L\, k\, \cos\!\left(\mathrm{\alpha_1}\right)\, \cos^2 \: \alpha_2 + 4\, L\, k\, \cos\!\left(\mathrm{\alpha_1}\right)\, \cos\!\left(\mathrm{\alpha_2}\right)\, \sin^2 \: \alpha_1 -  \\
    & \quad 4\, L\, k\, \cos\!\left(\mathrm{\alpha_1}\right)\, \cos\!\left(\mathrm{\alpha_2}\right)\, \sin\!\left(\mathrm{\alpha_1}\right)\, \sin\!\left(\mathrm{\alpha_2}\right) + 4\, L\, k\, \cos\!\left(\mathrm{\alpha_1}\right)\, \cos\!\left(\mathrm{\alpha_2}\right)\, \sin^2 \: \alpha_2 +  \\
    & \quad 4\, L\, k\, \cos\!\left(\mathrm{\alpha_1}\right)\, \cos\!\left(\mathrm{\alpha_2}\right) + 4\, L\, k\, \cos\!\left(\mathrm{\alpha_1}\right)\, \sin^2 \: \alpha_1\, \sin^2 \: \alpha_2 + 4\, L\, k\, \cos\!\left(\mathrm{\alpha_1}\right)\, \sin^2 \: \alpha_1 -  \\
    & \quad 4\, L\, k\, \cos\!\left(\mathrm{\alpha_1}\right)\, \sin\!\left(\mathrm{\alpha_1}\right)\, {\sin\!\left(\mathrm{\alpha_2}\right)}^3 - 4\, L\, k\, \cos\!\left(\mathrm{\alpha_1}\right)\, \sin\!\left(\mathrm{\alpha_1}\right)\, \sin\!\left(\mathrm{\alpha_2}\right) + 8\, L\, k\, \cos\!\left(\mathrm{\alpha_1}\right)\, {\sin\!\left(\mathrm{\alpha_2}\right)}^4 +  \\
    & \quad 12\, L\, k\, \cos\!\left(\mathrm{\alpha_1}\right)\, \sin^2 \: \alpha_2 + 4\, L\, k\, \cos\!\left(\mathrm{\alpha_1}\right) + 4\, L\, k\, \cos^4 \: \alpha_2\, \sin^2 \: \alpha_1 + 6\, L\, k\, \cos^4 \: \alpha_2 +  \\
    & \quad 4\, L\, k\, \cos^3 \: \alpha_2\, \sin^2 \: \alpha_1 + 4\, L\, k\, \cos^3 \: \alpha_2 + 8\, L\, k\, \cos^2 \: \alpha_2\, \sin^4 \: \alpha_1 +  \\
    & \quad 8\, L\, k\, \cos^2 \: \alpha_2\, \sin^2 \: \alpha_1\, \sin^2 \: \alpha_2 + 16\, L\, k\, \cos^2 \: \alpha_2\, \sin^2 \: \alpha_1 - 4\, L\, k\, \cos^2 \: \alpha_2\, \sin\!\left(\mathrm{\alpha_1}\right)\, \sin\!\left(\mathrm{\alpha_2}\right) +  \\
    & \quad 12\, L\, k\, \cos^2 \: \alpha_2\, \sin^2 \: \alpha_2 + 12\, L\, k\, \cos^2 \: \alpha_2 + 8\, L\, k\, \cos\!\left(\mathrm{\alpha_2}\right)\, \sin^4 \: \alpha_1 -  \\
    & \quad 4\, L\, k\, \cos\!\left(\mathrm{\alpha_2}\right)\, \sin^3 \: \alpha_1\, \sin\!\left(\mathrm{\alpha_2}\right) + 4\, L\, k\, \cos\!\left(\mathrm{\alpha_2}\right)\, \sin^2 \: \alpha_1\, \sin^2 \: \alpha_2 + 12\, L\, k\, \cos\!\left(\mathrm{\alpha_2}\right)\, \sin^2 \: \alpha_1 -  \\
    & \quad 4\, L\, k\, \cos\!\left(\mathrm{\alpha_2}\right)\, \sin\!\left(\mathrm{\alpha_1}\right)\, \sin\!\left(\mathrm{\alpha_2}\right) + 4\, L\, k\, \cos\!\left(\mathrm{\alpha_2}\right)\, \sin^2 \: \alpha_2 + 4\, L\, k\, \cos\!\left(\mathrm{\alpha_2}\right) + 4\, L\, k\, \sin^4 \: \alpha_1\, \sin^2 \: \alpha_2  \\
    & \quad + 6\, L\, k\, \sin^4 \: \alpha_1 - 4\, L\, k\, \sin^3 \: \alpha_1\, \sin\!\left(\mathrm{\alpha_2}\right) + 4\, L\, k\, \sin^2 \: \alpha_1\, {\sin\!\left(\mathrm{\alpha_2}\right)}^4 + 14\, L\, k\, \sin^2 \: \alpha_1\, \sin^2 \: \alpha_2 + \\
    & \quad  12\, L\, k\, \sin^2 \: \alpha_1 - 4\, L\, k\, \sin\!\left(\mathrm{\alpha_1}\right)\, {\sin\!\left(\mathrm{\alpha_2}\right)}^3 - 8\, L\, k\, \sin\!\left(\mathrm{\alpha_1}\right)\, \sin\!\left(\mathrm{\alpha_2}\right) + \\
    & \quad  6\, L\, k\, {\sin\!\left(\mathrm{\alpha_2}\right)}^4 + 12\, L\, k\, \sin^2 \: \alpha_2 + 6\, L\, k
\end{align*}

\begin{align*}
    N_1^{21} &= 2\, L^2\, k\, \sin\!\left(\mathrm{\alpha_2}\right)\, \left(4\, \cos^2 \: \alpha_1\, \cos^2 \: \alpha_2 + 4\, \cos^2 \: \alpha_1\, \cos\!\left(\mathrm{\alpha_2}\right) + 2\, \cos^2 \: \alpha_1\, \sin^2 \: \alpha_2 + \right.\\
    & \left. \quad  3\, \cos^2 \: \alpha_1 + 4\, \cos\!\left(\mathrm{\alpha_1}\right)\, \cos^2 \: \alpha_2 + 2\, \cos\!\left(\mathrm{\alpha_1}\right)\, \cos\!\left(\mathrm{\alpha_2}\right) + 2\, \cos\!\left(\mathrm{\alpha_1}\right)\, \sin^2 \: \alpha_2 + 2\, \cos\!\left(\mathrm{\alpha_1}\right) +  \right.\\
    & \left. \quad 2\, \cos^2 \: \alpha_2\, \sin^2 \: \alpha_1 + 3\, \cos^2 \: \alpha_2 + 2\, \cos\!\left(\mathrm{\alpha_2}\right)\, \sin^2 \: \alpha_1 + 2\, \cos\!\left(\mathrm{\alpha_2}\right) + \sin^2 \: \alpha_1\, \sin^2 \: \alpha_2 +  \right.\\
    & \left. \quad 2\, \sin^2 \: \alpha_1 + 2\, \sin^2 \: \alpha_2 + 3\right) \\
D_1^{21} &= D_2^{11}
\end{align*}
\begin{align*}
    N_2^{21} &= - \left(\frac{2\, L^3\, k}{3} + 2\, L^3\, k\, \left(\cos\!\left(\mathrm{\alpha_2}\right) + 1\right)\right)\, \left(2\, \cos^2 \: \alpha_1\, \cos\!\left(\mathrm{\alpha_2}\right)\, \sin\!\left(\mathrm{\alpha_2}\right) +  \cos^2 \: \alpha_1\, \sin\!\left(\mathrm{\alpha_1}\right) +  \right.\\
    & \left. \quad 2\, \cos^2 \: \alpha_1\, \sin\!\left(\mathrm{\alpha_2}\right) + 2\, \cos\!\left(\mathrm{\alpha_1}\right)\, \cos^2 \: \alpha_2\, \sin\!\left(\mathrm{\alpha_1}\right) +  \cos\!\left(\mathrm{\alpha_1}\right)\, \cos\!\left(\mathrm{\alpha_2}\right)\, \sin\!\left(\mathrm{\alpha_1}\right) +  \right.\\
    & \left. \quad  \cos\!\left(\mathrm{\alpha_1}\right)\, \cos\!\left(\mathrm{\alpha_2}\right)\, \sin\!\left(\mathrm{\alpha_2}\right) + \cos\!\left(\mathrm{\alpha_1}\right)\, \sin\!\left(\mathrm{\alpha_1}\right)\, \sin^2 \: \alpha_2 +  \cos\!\left(\mathrm{\alpha_1}\right)\, \sin\!\left(\mathrm{\alpha_1}\right) + 2\, \cos^2 \: \alpha_2\, \sin\!\left(\mathrm{\alpha_1}\right) +  \right.\\
    & \left. \quad  \cos^2 \: \alpha_2\, \sin\!\left(\mathrm{\alpha_2}\right) + \cos\!\left(\mathrm{\alpha_2}\right)\, \sin^2 \: \alpha_1\, \sin\!\left(\mathrm{\alpha_2}\right) + \cos\!\left(\mathrm{\alpha_2}\right)\, \sin\!\left(\mathrm{\alpha_2}\right) +  \right.\\
    & \left. \quad  \sin^3 \: \alpha_1 + \sin^2 \: \alpha_1\, \sin\!\left(\mathrm{\alpha_2}\right) + \sin\!\left(\mathrm{\alpha_1}\right)\, \sin^2 \: \alpha_2 + 2\, \sin\!\left(\mathrm{\alpha_1}\right) + {\sin\!\left(\mathrm{\alpha_2}\right)}^3 + 2\, \sin\!\left(\mathrm{\alpha_2}\right)\right)
\end{align*}
\begin{align*}
    D_2^{21} &= 8\, k\, L^2\, \cos^4 \: \alpha_1\, \cos^2 \: \alpha_2 + 8\, k\, L^2\, \cos^4 \: \alpha_1\, \cos\!\left(\mathrm{\alpha_2}\right) + 4\, k\, L^2\, \cos^4 \: \alpha_1\, \sin^2 \: \alpha_2 + 6\, k\, L^2\, \cos^4 \: \alpha_1 + \\
    & \quad 8\, k\, L^2\, \cos^3 \: \alpha_1\, \cos^2 \: \alpha_2 + 4\, k\, L^2\, \cos^3 \: \alpha_1\, \cos\!\left(\mathrm{\alpha_2}\right) + 4\, k\, L^2\, \cos^3 \: \alpha_1\, \sin^2 \: \alpha_2 +  \\
    & \quad 4\, k\, L^2\, \cos^3 \: \alpha_1 + 8\, k\, L^2\, \cos^2 \: \alpha_1\, \cos^4 \: \alpha_2 + 8\, k\, L^2\, \cos^2 \: \alpha_1\, \cos^3 \: \alpha_2 +  \\
    & \quad 16\, k\, L^2\, \cos^2 \: \alpha_1\, \cos^2 \: \alpha_2\, \sin^2 \: \alpha_1 + 16\, k\, L^2\, \cos^2 \: \alpha_1\, \cos^2 \: \alpha_2\, \sin^2 \: \alpha_2 +  \\
    & \quad 20\, k\, L^2\, \cos^2 \: \alpha_1\, \cos^2 \: \alpha_2 + 16\, k\, L^2\, \cos^2 \: \alpha_1\, \cos\!\left(\mathrm{\alpha_2}\right)\, \sin^2 \: \alpha_1 -  \\
    & \quad 4\, k\, L^2\, \cos^2 \: \alpha_1\, \cos\!\left(\mathrm{\alpha_2}\right)\, \sin\!\left(\mathrm{\alpha_1}\right)\, \sin\!\left(\mathrm{\alpha_2}\right) + 8\, k\, L^2\, \cos^2 \: \alpha_1\, \cos\!\left(\mathrm{\alpha_2}\right)\, \sin^2 \: \alpha_2 +  \\
    & \quad 12\, k\, L^2\, \cos^2 \: \alpha_1\, \cos\!\left(\mathrm{\alpha_2}\right) + 8\, k\, L^2\, \cos^2 \: \alpha_1\, \sin^2 \: \alpha_1\, \sin^2 \: \alpha_2 + 12\, k\, L^2\, \cos^2 \: \alpha_1\, \sin^2 \: \alpha_1 -  \\
    & \quad 4\, k\, L^2\, \cos^2 \: \alpha_1\, \sin\!\left(\mathrm{\alpha_1}\right)\, \sin\!\left(\mathrm{\alpha_2}\right) + 8\, k\, L^2\, \cos^2 \: \alpha_1\, {\sin\!\left(\mathrm{\alpha_2}\right)}^4 + 16\, k\, L^2\, \cos^2 \: \alpha_1\, \sin^2 \: \alpha_2 +  \\
    & \quad 12\, k\, L^2\, \cos^2 \: \alpha_1 + 8\, k\, L^2\, \cos\!\left(\mathrm{\alpha_1}\right)\, \cos^4 \: \alpha_2 + 4\, k\, L^2\, \cos\!\left(\mathrm{\alpha_1}\right)\, \cos^3 \: \alpha_2 +  \\
    & \quad 8\, k\, L^2\, \cos\!\left(\mathrm{\alpha_1}\right)\, \cos^2 \: \alpha_2\, \sin^2 \: \alpha_1 - 4\, k\, L^2\, \cos\!\left(\mathrm{\alpha_1}\right)\, \cos^2 \: \alpha_2\, \sin\!\left(\mathrm{\alpha_1}\right)\, \sin\!\left(\mathrm{\alpha_2}\right) +  \\
    & \quad 16\, k\, L^2\, \cos\!\left(\mathrm{\alpha_1}\right)\, \cos^2 \: \alpha_2\, \sin^2 \: \alpha_2 + 12\, k\, L^2\, \cos\!\left(\mathrm{\alpha_1}\right)\, \cos^2 \: \alpha_2 + 4\, k\, L^2\, \cos\!\left(\mathrm{\alpha_1}\right)\, \cos\!\left(\mathrm{\alpha_2}\right)\, \sin^2 \: \alpha_1 -  \\
    & \quad 4\, k\, L^2\, \cos\!\left(\mathrm{\alpha_1}\right)\, \cos\!\left(\mathrm{\alpha_2}\right)\, \sin\!\left(\mathrm{\alpha_1}\right)\, \sin\!\left(\mathrm{\alpha_2}\right) + 4\, k\, L^2\, \cos\!\left(\mathrm{\alpha_1}\right)\, \cos\!\left(\mathrm{\alpha_2}\right)\, \sin^2 \: \alpha_2 +  \\
    & \quad 4\, k\, L^2\, \cos\!\left(\mathrm{\alpha_1}\right)\, \cos\!\left(\mathrm{\alpha_2}\right) + 4\, k\, L^2\, \cos\!\left(\mathrm{\alpha_1}\right)\, \sin^2 \: \alpha_1\, \sin^2 \: \alpha_2 + 4\, k\, L^2\, \cos\!\left(\mathrm{\alpha_1}\right)\, \sin^2 \: \alpha_1 -  \\
    & \quad 4\, k\, L^2\, \cos\!\left(\mathrm{\alpha_1}\right)\, \sin\!\left(\mathrm{\alpha_1}\right)\, {\sin\!\left(\mathrm{\alpha_2}\right)}^3 - 4\, k\, L^2\, \cos\!\left(\mathrm{\alpha_1}\right)\, \sin\!\left(\mathrm{\alpha_1}\right)\, \sin\!\left(\mathrm{\alpha_2}\right) + 8\, k\, L^2\, \cos\!\left(\mathrm{\alpha_1}\right)\, {\sin\!\left(\mathrm{\alpha_2}\right)}^4 +  \\
    & \quad 12\, k\, L^2\, \cos\!\left(\mathrm{\alpha_1}\right)\, \sin^2 \: \alpha_2 + 4\, k\, L^2\, \cos\!\left(\mathrm{\alpha_1}\right) + 4\, k\, L^2\, \cos^4 \: \alpha_2\, \sin^2 \: \alpha_1 + 6\, k\, L^2\, \cos^4 \: \alpha_2 +  \\
    & \quad 4\, k\, L^2\, \cos^3 \: \alpha_2\, \sin^2 \: \alpha_1 + 4\, k\, L^2\, \cos^3 \: \alpha_2 + 8\, k\, L^2\, \cos^2 \: \alpha_2\, \sin^4 \: \alpha_1 +  \\
    & \quad 8\, k\, L^2\, \cos^2 \: \alpha_2\, \sin^2 \: \alpha_1\, \sin^2 \: \alpha_2 + 16\, k\, L^2\, \cos^2 \: \alpha_2\, \sin^2 \: \alpha_1 - 4\, k\, L^2\, \cos^2 \: \alpha_2\, \sin\!\left(\mathrm{\alpha_1}\right)\, \sin\!\left(\mathrm{\alpha_2}\right) +  \\
    & \quad 12\, k\, L^2\, \cos^2 \: \alpha_2\, \sin^2 \: \alpha_2 + 12\, k\, L^2\, \cos^2 \: \alpha_2 + 8\, k\, L^2\, \cos\!\left(\mathrm{\alpha_2}\right)\, \sin^4 \: \alpha_1 -  \\
    & \quad 4\, k\, L^2\, \cos\!\left(\mathrm{\alpha_2}\right)\, \sin^3 \: \alpha_1\, \sin\!\left(\mathrm{\alpha_2}\right) + 4\, k\, L^2\, \cos\!\left(\mathrm{\alpha_2}\right)\, \sin^2 \: \alpha_1\, \sin^2 \: \alpha_2 + 12\, k\, L^2\, \cos\!\left(\mathrm{\alpha_2}\right)\, \sin^2 \: \alpha_1 -  \\
    & \quad 4\, k\, L^2\, \cos\!\left(\mathrm{\alpha_2}\right)\, \sin\!\left(\mathrm{\alpha_1}\right)\, \sin\!\left(\mathrm{\alpha_2}\right) + 4\, k\, L^2\, \cos\!\left(\mathrm{\alpha_2}\right)\, \sin^2 \: \alpha_2 +  \\
    & \quad 4\, k\, L^2\, \cos\!\left(\mathrm{\alpha_2}\right) + 4\, k\, L^2\, \sin^4 \: \alpha_1\, \sin^2 \: \alpha_2 + 6\, k\, L^2\, \sin^4 \: \alpha_1 - 4\, k\, L^2\, \sin^3 \: \alpha_1\, \sin\!\left(\mathrm{\alpha_2}\right) +  \\
    & \quad 4\, k\, L^2\, \sin^2 \: \alpha_1\, {\sin\!\left(\mathrm{\alpha_2}\right)}^4 + 14\, k\, L^2\, \sin^2 \: \alpha_1\, \sin^2 \: \alpha_2 + 12\, k\, L^2\, \sin^2 \: \alpha_1 - 4\, k\, L^2\, \sin\!\left(\mathrm{\alpha_1}\right)\, {\sin\!\left(\mathrm{\alpha_2}\right)}^3 -  \\
    & \quad 8\, k\, L^2\, \sin\!\left(\mathrm{\alpha_1}\right)\, \sin\!\left(\mathrm{\alpha_2}\right) + 6\, k\, L^2\, {\sin\!\left(\mathrm{\alpha_2}\right)}^4 + 12\, k\, L^2\, \sin^2 \: \alpha_2 + 6\, k\, L^2
\end{align*}
\begin{align*}
    N_3^{21} &= - L^2\, k\, \cos\!\left(\mathrm{\alpha_2}\right)\, \left(2\, \cos^2 \: \alpha_1\, \cos\!\left(\mathrm{\alpha_2}\right)\, \sin\!\left(\mathrm{\alpha_2}\right) + \cos^2 \: \alpha_1\, \sin\!\left(\mathrm{\alpha_1}\right) + 2\, \cos^2 \: \alpha_1\, \sin\!\left(\mathrm{\alpha_2}\right) -  \right.\\
    & \left. \quad 2\, \cos\!\left(\mathrm{\alpha_1}\right)\, \cos^2 \: \alpha_2\, \sin\!\left(\mathrm{\alpha_1}\right) - 3\, \cos\!\left(\mathrm{\alpha_1}\right)\, \cos\!\left(\mathrm{\alpha_2}\right)\, \sin\!\left(\mathrm{\alpha_1}\right) + 3\, \cos\!\left(\mathrm{\alpha_1}\right)\, \cos\!\left(\mathrm{\alpha_2}\right)\, \sin\!\left(\mathrm{\alpha_2}\right) -  \right.\\
    & \left. \quad  \cos\!\left(\mathrm{\alpha_1}\right)\, \sin\!\left(\mathrm{\alpha_1}\right)\, \sin^2 \: \alpha_2 - \cos\!\left(\mathrm{\alpha_1}\right)\, \sin\!\left(\mathrm{\alpha_1}\right) + 2\, \cos\!\left(\mathrm{\alpha_1}\right)\, \sin\!\left(\mathrm{\alpha_2}\right) - 2\, \cos^2 \: \alpha_2\, \sin\!\left(\mathrm{\alpha_1}\right) -  \right.\\
    & \left. \quad  \cos^2 \: \alpha_2\, \sin\!\left(\mathrm{\alpha_2}\right) + \cos\!\left(\mathrm{\alpha_2}\right)\, \sin^2 \: \alpha_1\, \sin\!\left(\mathrm{\alpha_2}\right) - 2\, \cos\!\left(\mathrm{\alpha_2}\right)\, \sin\!\left(\mathrm{\alpha_1}\right) + \right.\\
    & \left. \quad \cos\!\left(\mathrm{\alpha_2}\right)\, \sin\!\left(\mathrm{\alpha_2}\right) + \sin^3 \: \alpha_1 + \sin^2 \: \alpha_1\, \sin\!\left(\mathrm{\alpha_2}\right) - \sin\!\left(\mathrm{\alpha_1}\right)\, \sin^2 \: \alpha_2 - {\sin\!\left(\mathrm{\alpha_2}\right)}^3\right) \\
D_3^{21} &= D_2^{11}
\end{align*}
\begin{align*}
    N_1^{22} &= L^2\, k\, \sin\!\left(\mathrm{\alpha_2}\right)\, \left(2\, \cos^2 \: \alpha_1\, \cos\!\left(\mathrm{\alpha_2}\right)\, \sin\!\left(\mathrm{\alpha_2}\right) + \cos^2 \: \alpha_1\, \sin\!\left(\mathrm{\alpha_1}\right) + 2\, \cos^2 \: \alpha_1\, \sin\!\left(\mathrm{\alpha_2}\right) - \right.\\
    & \left. \quad 2\, \cos\!\left(\mathrm{\alpha_1}\right)\, \cos^2 \: \alpha_2\, \sin\!\left(\mathrm{\alpha_1}\right) - 3\, \cos\!\left(\mathrm{\alpha_1}\right)\, \cos\!\left(\mathrm{\alpha_2}\right)\, \sin\!\left(\mathrm{\alpha_1}\right) + 3\, \cos\!\left(\mathrm{\alpha_1}\right)\, \cos\!\left(\mathrm{\alpha_2}\right)\, \sin\!\left(\mathrm{\alpha_2}\right) -\right.\\
    & \left. \quad  \cos\!\left(\mathrm{\alpha_1}\right)\, \sin\!\left(\mathrm{\alpha_1}\right)\, \sin^2 \: \alpha_2 - \cos\!\left(\mathrm{\alpha_1}\right)\, \sin\!\left(\mathrm{\alpha_1}\right) + 2\, \cos\!\left(\mathrm{\alpha_1}\right)\, \sin\!\left(\mathrm{\alpha_2}\right) - 2\, \cos^2 \: \alpha_2\, \sin\!\left(\mathrm{\alpha_1}\right) -\right.\\
    & \left. \quad  \cos^2 \: \alpha_2\, \sin\!\left(\mathrm{\alpha_2}\right) + \cos\!\left(\mathrm{\alpha_2}\right)\, \sin^2 \: \alpha_1\, \sin\!\left(\mathrm{\alpha_2}\right) - 2\, \cos\!\left(\mathrm{\alpha_2}\right)\, \sin\!\left(\mathrm{\alpha_1}\right) +\right.\\
    & \left. \quad  \cos\!\left(\mathrm{\alpha_2}\right)\, \sin\!\left(\mathrm{\alpha_2}\right) + \sin^3 \: \alpha_1 + \sin^2 \: \alpha_1\, \sin\!\left(\mathrm{\alpha_2}\right) - \sin\!\left(\mathrm{\alpha_1}\right)\, \sin^2 \: \alpha_2 - {\sin\!\left(\mathrm{\alpha_2}\right)}^3\right) \\
    D_1^{22} &= D_2^{11}\\
\end{align*}
\begin{align*}
    N_2^{22} &= - \left(\frac{2\, L^3\, k}{3} + 2\, L^3\, k\, \left(\cos\!\left(\mathrm{\alpha_2}\right) + 1\right)\right)\, \left(2\, \cos^4 \: \alpha_1 + 2\, \cos^3 \: \alpha_1 - 2\, \cos^2 \: \alpha_1\, \cos\!\left(\mathrm{\alpha_2}\right) + \right.\\
    & \left. \quad 4\, \cos^2 \: \alpha_1\, \sin^2 \: \alpha_1 + 3\, \cos^2 \: \alpha_1\, \sin^2 \: \alpha_2 + 2\, \cos^2 \: \alpha_1 + 2\, \cos\!\left(\mathrm{\alpha_1}\right)\, \cos^2 \: \alpha_2 + \right.\\
    & \left. \quad 2\, \cos\!\left(\mathrm{\alpha_1}\right)\, \sin^2 \: \alpha_1 - 2\, \cos\!\left(\mathrm{\alpha_1}\right)\, \sin\!\left(\mathrm{\alpha_1}\right)\, \sin\!\left(\mathrm{\alpha_2}\right) + 4\, \cos\!\left(\mathrm{\alpha_1}\right)\, \sin^2 \: \alpha_2 + 2\, \cos\!\left(\mathrm{\alpha_1}\right) - \right.\\
    & \left. \quad 2\, \cos^4 \: \alpha_2 - 2\, \cos^3 \: \alpha_2 - 3\, \cos^2 \: \alpha_2\, \sin^2 \: \alpha_1 - 4\, \cos^2 \: \alpha_2\, \sin^2 \: \alpha_2 - 2\, \cos^2 \: \alpha_2 - \right.\\
    & \left. \quad 4\, \cos\!\left(\mathrm{\alpha_2}\right)\, \sin^2 \: \alpha_1 + 2\, \cos\!\left(\mathrm{\alpha_2}\right)\, \sin\!\left(\mathrm{\alpha_1}\right)\, \sin\!\left(\mathrm{\alpha_2}\right) - 2\, \cos\!\left(\mathrm{\alpha_2}\right)\, \sin^2 \: \alpha_2 - 2\, \cos\!\left(\mathrm{\alpha_2}\right) + \right.\\
    & \left. \quad 2\, \sin^4 \: \alpha_1 + \sin^2 \: \alpha_1 - 2\, {\sin\!\left(\mathrm{\alpha_2}\right)}^4 - \sin^2 \: \alpha_2\right) \\
    D_2^{22} &= D_1^{12}\\
\end{align*}
\begin{align*}
    N_3^{22} &= - L^2\, k\, \cos\!\left(\mathrm{\alpha_2}\right)\, \left(2\, \cos^4 \: \alpha_1 + 4\, \cos^3 \: \alpha_1 + 4\, \cos^2 \: \alpha_1\, \cos^2 \: \alpha_2 + 4\, \cos^2 \: \alpha_1\, \cos\!\left(\mathrm{\alpha_2}\right) + \right.\\
    & \left. \quad 4\, \cos^2 \: \alpha_1\, \sin^2 \: \alpha_1 + 5\, \cos^2 \: \alpha_1\, \sin^2 \: \alpha_2 + 8\, \cos^2 \: \alpha_1 + 4\, \cos\!\left(\mathrm{\alpha_1}\right)\, \cos^2 \: \alpha_2 - \right.\\
    & \left. \quad 2\, \cos\!\left(\mathrm{\alpha_1}\right)\, \cos\!\left(\mathrm{\alpha_2}\right)\, \sin\!\left(\mathrm{\alpha_1}\right)\, \sin\!\left(\mathrm{\alpha_2}\right) + 4\, \cos\!\left(\mathrm{\alpha_1}\right)\, \sin^2 \: \alpha_1 - 4\, \cos\!\left(\mathrm{\alpha_1}\right)\, \sin\!\left(\mathrm{\alpha_1}\right)\, \sin\!\left(\mathrm{\alpha_2}\right) + \right.\\
    & \left. \quad 8\, \cos\!\left(\mathrm{\alpha_1}\right)\, \sin^2 \: \alpha_2 + 4\, \cos\!\left(\mathrm{\alpha_1}\right) + 2\, \cos^4 \: \alpha_2 + \right.\\
    & \left. \quad 4\, \cos^3 \: \alpha_2 + 5\, \cos^2 \: \alpha_2\, \sin^2 \: \alpha_1 + 4\, \cos^2 \: \alpha_2\, \sin^2 \: \alpha_2 + 8\, \cos^2 \: \alpha_2 + 8\, \cos\!\left(\mathrm{\alpha_2}\right)\, \sin^2 \: \alpha_1 - \right.\\
    & \left. \quad 4\, \cos\!\left(\mathrm{\alpha_2}\right)\, \sin\!\left(\mathrm{\alpha_1}\right)\, \sin\!\left(\mathrm{\alpha_2}\right) + 4\, \cos\!\left(\mathrm{\alpha_2}\right)\, \sin^2 \: \alpha_2 + 4\, \cos\!\left(\mathrm{\alpha_2}\right) + 2\, \sin^4 \: \alpha_1 + 4\, \sin^2 \: \alpha_1\, \sin^2 \: \alpha_2 \right.\\
    & \left. \quad + 9\, \sin^2 \: \alpha_1 - 8\, \sin\!\left(\mathrm{\alpha_1}\right)\, \sin\!\left(\mathrm{\alpha_2}\right) + 2\, {\sin\!\left(\mathrm{\alpha_2}\right)}^4 + 9\, \sin^2 \: \alpha_2 + 6\right) \\
D_3^{22} &= D_3^{12}\\
\end{align*}

\begin{align*}
    N_1^{31} &= \left(\frac{2\, L^3\, k}{3} + 2\, L^3\, k\, \left(\cos\!\left(\mathrm{\alpha_1}\right) + 1\right)\right)\, \left(2\, \cos^4 \: \alpha_1 + 4\, \cos^2 \: \alpha_1\, \cos^2 \: \alpha_2 + 4\, \cos^2 \: \alpha_1\, \sin^2 \: \alpha_1 + \right.\\
    & \left. \quad 5\, \cos^2 \: \alpha_1\, \sin^2 \: \alpha_2 + 4\, \cos^2 \: \alpha_1 + 2\, \cos\!\left(\mathrm{\alpha_1}\right)\, \cos\!\left(\mathrm{\alpha_2}\right)\, \sin\!\left(\mathrm{\alpha_1}\right)\, \sin\!\left(\mathrm{\alpha_2}\right) + 2\, \cos^4 \: \alpha_2 + \right.\\
    & \left. \quad 5\, \cos^2 \: \alpha_2\, \sin^2 \: \alpha_1 + 4\, \cos^2 \: \alpha_2\, \sin^2 \: \alpha_2 + 4\, \cos^2 \: \alpha_2 + 2\, \sin^4 \: \alpha_1 + 4\, \sin^2 \: \alpha_1\, \sin^2 \: \alpha_2 + \right.\\
    & \left. \quad 5\, \sin^2 \: \alpha_1 + 2\, {\sin\!\left(\mathrm{\alpha_2}\right)}^4 + 5\, \sin^2 \: \alpha_2 + 2\right)
\\
\end{align*}
\begin{align*}
    D_1^{31} &= 16\, k\, L^3\, \cos^4 \: \alpha_1\, \cos^2 \: \alpha_2 + 16\, k\, L^3\, \cos^4 \: \alpha_1\, \cos\!\left(\mathrm{\alpha_2}\right) + 8\, k\, L^3\, \cos^4 \: \alpha_1\, \sin^2 \: \alpha_2 + \\
    & \quad 12\, k\, L^3\, \cos^4 \: \alpha_1 + 16\, k\, L^3\, \cos^3 \: \alpha_1\, \cos^2 \: \alpha_2 + 8\, k\, L^3\, \cos^3 \: \alpha_1\, \cos\!\left(\mathrm{\alpha_2}\right) + 8\, k\, L^3\, \cos^3 \: \alpha_1\, \sin^2 \: \alpha_2 + \\
    & \quad 8\, k\, L^3\, \cos^3 \: \alpha_1 + 16\, k\, L^3\, \cos^2 \: \alpha_1\, \cos^4 \: \alpha_2 + 16\, k\, L^3\, \cos^2 \: \alpha_1\, \cos^3 \: \alpha_2 + \\
    & \quad 32\, k\, L^3\, \cos^2 \: \alpha_1\, \cos^2 \: \alpha_2\, \sin^2 \: \alpha_1 + 32\, k\, L^3\, \cos^2 \: \alpha_1\, \cos^2 \: \alpha_2\, \sin^2 \: \alpha_2 + 40\, k\, L^3\, \cos^2 \: \alpha_1\, \cos^2 \: \alpha_2 \\
    & \quad + 32\, k\, L^3\, \cos^2 \: \alpha_1\, \cos\!\left(\mathrm{\alpha_2}\right)\, \sin^2 \: \alpha_1 - 8\, k\, L^3\, \cos^2 \: \alpha_1\, \cos\!\left(\mathrm{\alpha_2}\right)\, \sin\!\left(\mathrm{\alpha_1}\right)\, \sin\!\left(\mathrm{\alpha_2}\right) + \\
    & \quad 16\, k\, L^3\, \cos^2 \: \alpha_1\, \cos\!\left(\mathrm{\alpha_2}\right)\, \sin^2 \: \alpha_2 + 24\, k\, L^3\, \cos^2 \: \alpha_1\, \cos\!\left(\mathrm{\alpha_2}\right) + 16\, k\, L^3\, \cos^2 \: \alpha_1\, \sin^2 \: \alpha_1\, \sin^2 \: \alpha_2 \\
    & \quad + 24\, k\, L^3\, \cos^2 \: \alpha_1\, \sin^2 \: \alpha_1 - 8\, k\, L^3\, \cos^2 \: \alpha_1\, \sin\!\left(\mathrm{\alpha_1}\right)\, \sin\!\left(\mathrm{\alpha_2}\right) + 16\, k\, L^3\, \cos^2 \: \alpha_1\, {\sin\!\left(\mathrm{\alpha_2}\right)}^4 +  \\
    & \quad 32\, k\, L^3\, \cos^2 \: \alpha_1\, \sin^2 \: \alpha_2 + 24\, k\, L^3\, \cos^2 \: \alpha_1 + 16\, k\, L^3\, \cos\!\left(\mathrm{\alpha_1}\right)\, \cos^4 \: \alpha_2 +  \\
    & \quad 8\, k\, L^3\, \cos\!\left(\mathrm{\alpha_1}\right)\, \cos^3 \: \alpha_2 + 16\, k\, L^3\, \cos\!\left(\mathrm{\alpha_1}\right)\, \cos^2 \: \alpha_2\, \sin^2 \: \alpha_1 -  \\
    & \quad 8\, k\, L^3\, \cos\!\left(\mathrm{\alpha_1}\right)\, \cos^2 \: \alpha_2\, \sin\!\left(\mathrm{\alpha_1}\right)\, \sin\!\left(\mathrm{\alpha_2}\right) + 32\, k\, L^3\, \cos\!\left(\mathrm{\alpha_1}\right)\, \cos^2 \: \alpha_2\, \sin^2 \: \alpha_2 +  \\
    & \quad 24\, k\, L^3\, \cos\!\left(\mathrm{\alpha_1}\right)\, \cos^2 \: \alpha_2 + 8\, k\, L^3\, \cos\!\left(\mathrm{\alpha_1}\right)\, \cos\!\left(\mathrm{\alpha_2}\right)\, \sin^2 \: \alpha_1 -  \\
    & \quad 8\, k\, L^3\, \cos\!\left(\mathrm{\alpha_1}\right)\, \cos\!\left(\mathrm{\alpha_2}\right)\, \sin\!\left(\mathrm{\alpha_1}\right)\, \sin\!\left(\mathrm{\alpha_2}\right) + 8\, k\, L^3\, \cos\!\left(\mathrm{\alpha_1}\right)\, \cos\!\left(\mathrm{\alpha_2}\right)\, \sin^2 \: \alpha_2 +  \\
    & \quad 8\, k\, L^3\, \cos\!\left(\mathrm{\alpha_1}\right)\, \cos\!\left(\mathrm{\alpha_2}\right) + 8\, k\, L^3\, \cos\!\left(\mathrm{\alpha_1}\right)\, \sin^2 \: \alpha_1\, \sin^2 \: \alpha_2 + 8\, k\, L^3\, \cos\!\left(\mathrm{\alpha_1}\right)\, \sin^2 \: \alpha_1 -  \\
    & \quad 8\, k\, L^3\, \cos\!\left(\mathrm{\alpha_1}\right)\, \sin\!\left(\mathrm{\alpha_1}\right)\, {\sin\!\left(\mathrm{\alpha_2}\right)}^3 - 8\, k\, L^3\, \cos\!\left(\mathrm{\alpha_1}\right)\, \sin\!\left(\mathrm{\alpha_1}\right)\, \sin\!\left(\mathrm{\alpha_2}\right) + 16\, k\, L^3\, \cos\!\left(\mathrm{\alpha_1}\right)\, {\sin\!\left(\mathrm{\alpha_2}\right)}^4 +  \\
    & \quad 24\, k\, L^3\, \cos\!\left(\mathrm{\alpha_1}\right)\, \sin^2 \: \alpha_2 + 8\, k\, L^3\, \cos\!\left(\mathrm{\alpha_1}\right) + 8\, k\, L^3\, \cos^4 \: \alpha_2\, \sin^2 \: \alpha_1 + 12\, k\, L^3\, \cos^4 \: \alpha_2 +  \\
    & \quad 8\, k\, L^3\, \cos^3 \: \alpha_2\, \sin^2 \: \alpha_1 + 8\, k\, L^3\, \cos^3 \: \alpha_2 + 16\, k\, L^3\, \cos^2 \: \alpha_2\, \sin^4 \: \alpha_1 +  \\
    & \quad 16\, k\, L^3\, \cos^2 \: \alpha_2\, \sin^2 \: \alpha_1\, \sin^2 \: \alpha_2 + 32\, k\, L^3\, \cos^2 \: \alpha_2\, \sin^2 \: \alpha_1 - 8\, k\, L^3\, \cos^2 \: \alpha_2\, \sin\!\left(\mathrm{\alpha_1}\right)\, \sin\!\left(\mathrm{\alpha_2}\right) +  \\
    & \quad 24\, k\, L^3\, \cos^2 \: \alpha_2\, \sin^2 \: \alpha_2 + 24\, k\, L^3\, \cos^2 \: \alpha_2 + 16\, k\, L^3\, \cos\!\left(\mathrm{\alpha_2}\right)\, \sin^4 \: \alpha_1 -  \\
    & \quad 8\, k\, L^3\, \cos\!\left(\mathrm{\alpha_2}\right)\, \sin^3 \: \alpha_1\, \sin\!\left(\mathrm{\alpha_2}\right) + 8\, k\, L^3\, \cos\!\left(\mathrm{\alpha_2}\right)\, \sin^2 \: \alpha_1\, \sin^2 \: \alpha_2 + 24\, k\, L^3\, \cos\!\left(\mathrm{\alpha_2}\right)\, \sin^2 \: \alpha_1 -  \\
    & \quad 8\, k\, L^3\, \cos\!\left(\mathrm{\alpha_2}\right)\, \sin\!\left(\mathrm{\alpha_1}\right)\, \sin\!\left(\mathrm{\alpha_2}\right) + 8\, k\, L^3\, \cos\!\left(\mathrm{\alpha_2}\right)\, \sin^2 \: \alpha_2 + 8\, k\, L^3\, \cos\!\left(\mathrm{\alpha_2}\right) +  \\
    & \quad 8\, k\, L^3\, \sin^4 \: \alpha_1\, \sin^2 \: \alpha_2 + 12\, k\, L^3\, \sin^4 \: \alpha_1 - 8\, k\, L^3\, \sin^3 \: \alpha_1\, \sin\!\left(\mathrm{\alpha_2}\right) + 8\, k\, L^3\, \sin^2 \: \alpha_1\, {\sin\!\left(\mathrm{\alpha_2}\right)}^4 +  \\
    & \quad 28\, k\, L^3\, \sin^2 \: \alpha_1\, \sin^2 \: \alpha_2 + 24\, k\, L^3\, \sin^2 \: \alpha_1 - 8\, k\, L^3\, \sin\!\left(\mathrm{\alpha_1}\right)\, {\sin\!\left(\mathrm{\alpha_2}\right)}^3 - 16\, k\, L^3\, \sin\!\left(\mathrm{\alpha_1}\right)\, \sin\!\left(\mathrm{\alpha_2}\right) +  \\
    & \quad 12\, k\, L^3\, {\sin\!\left(\mathrm{\alpha_2}\right)}^4 + 24\, k\, L^3\, \sin^2 \: \alpha_2 + 12\, k\, L^3
\end{align*}
\begin{align*}
    N_2^{31} &= - L^2\, k\, \sin\!\left(\mathrm{\alpha_1}\right)\, \left(2\, \cos^2 \: \alpha_1\, \cos\!\left(\mathrm{\alpha_2}\right)\, \sin\!\left(\mathrm{\alpha_2}\right) + \cos^2 \: \alpha_1\, \sin\!\left(\mathrm{\alpha_1}\right) + 2\, \cos^2 \: \alpha_1\, \sin\!\left(\mathrm{\alpha_2}\right) + \right.\\
    & \left. \quad 2\, \cos\!\left(\mathrm{\alpha_1}\right)\, \cos^2 \: \alpha_2\, \sin\!\left(\mathrm{\alpha_1}\right) + \cos\!\left(\mathrm{\alpha_1}\right)\, \cos\!\left(\mathrm{\alpha_2}\right)\, \sin\!\left(\mathrm{\alpha_1}\right) + \cos\!\left(\mathrm{\alpha_1}\right)\, \cos\!\left(\mathrm{\alpha_2}\right)\, \sin\!\left(\mathrm{\alpha_2}\right) +\right.\\
    & \left. \quad  \cos\!\left(\mathrm{\alpha_1}\right)\, \sin\!\left(\mathrm{\alpha_1}\right)\, \sin^2 \: \alpha_2 + \cos\!\left(\mathrm{\alpha_1}\right)\, \sin\!\left(\mathrm{\alpha_1}\right) + 2\, \cos^2 \: \alpha_2\, \sin\!\left(\mathrm{\alpha_1}\right) + \cos^2 \: \alpha_2\, \sin\!\left(\mathrm{\alpha_2}\right) +\right.\\
    & \left. \quad  \cos\!\left(\mathrm{\alpha_2}\right)\, \sin^2 \: \alpha_1\, \sin\!\left(\mathrm{\alpha_2}\right) + \cos\!\left(\mathrm{\alpha_2}\right)\, \sin\!\left(\mathrm{\alpha_2}\right) + \sin^3 \: \alpha_1 +\right.\\
    & \left. \quad \sin^2 \: \alpha_1\, \sin\!\left(\mathrm{\alpha_2}\right) + \sin\!\left(\mathrm{\alpha_1}\right)\, \sin^2 \: \alpha_2 + 2\, \sin\!\left(\mathrm{\alpha_1}\right) + {\sin\!\left(\mathrm{\alpha_2}\right)}^3 + 2\, \sin\!\left(\mathrm{\alpha_2}\right)\right)
\end{align*}
\begin{align*}
    D_2^{31} &= 4\, k\, L^2\, \cos^4 \: \alpha_1\, \cos^2 \: \alpha_2 + 4\, k\, L^2\, \cos^4 \: \alpha_1\, \cos\!\left(\mathrm{\alpha_2}\right) + 2\, k\, L^2\, \cos^4 \: \alpha_1\, \sin^2 \: \alpha_2 + 3\, k\, L^2\, \cos^4 \: \alpha_1 + \\
    & \quad 4\, k\, L^2\, \cos^3 \: \alpha_1\, \cos^2 \: \alpha_2 + 2\, k\, L^2\, \cos^3 \: \alpha_1\, \cos\!\left(\mathrm{\alpha_2}\right) + 2\, k\, L^2\, \cos^3 \: \alpha_1\, \sin^2 \: \alpha_2 + 2\, k\, L^2\, \cos^3 \: \alpha_1 + \\
    & \quad 4\, k\, L^2\, \cos^2 \: \alpha_1\, \cos^4 \: \alpha_2 + 4\, k\, L^2\, \cos^2 \: \alpha_1\, \cos^3 \: \alpha_2 + 8\, k\, L^2\, \cos^2 \: \alpha_1\, \cos^2 \: \alpha_2\, \sin^2 \: \alpha_1 + \\
    & \quad 8\, k\, L^2\, \cos^2 \: \alpha_1\, \cos^2 \: \alpha_2\, \sin^2 \: \alpha_2 + 10\, k\, L^2\, \cos^2 \: \alpha_1\, \cos^2 \: \alpha_2 + 8\, k\, L^2\, \cos^2 \: \alpha_1\, \cos\!\left(\mathrm{\alpha_2}\right)\, \sin^2 \: \alpha_1 - \\
    & \quad 2\, k\, L^2\, \cos^2 \: \alpha_1\, \cos\!\left(\mathrm{\alpha_2}\right)\, \sin\!\left(\mathrm{\alpha_1}\right)\, \sin\!\left(\mathrm{\alpha_2}\right) + 4\, k\, L^2\, \cos^2 \: \alpha_1\, \cos\!\left(\mathrm{\alpha_2}\right)\, \sin^2 \: \alpha_2 + \\
    & \quad 6\, k\, L^2\, \cos^2 \: \alpha_1\, \cos\!\left(\mathrm{\alpha_2}\right) + 4\, k\, L^2\, \cos^2 \: \alpha_1\, \sin^2 \: \alpha_1\, \sin^2 \: \alpha_2 + 6\, k\, L^2\, \cos^2 \: \alpha_1\, \sin^2 \: \alpha_1 - \\
    & \quad 2\, k\, L^2\, \cos^2 \: \alpha_1\, \sin\!\left(\mathrm{\alpha_1}\right)\, \sin\!\left(\mathrm{\alpha_2}\right) + 4\, k\, L^2\, \cos^2 \: \alpha_1\, {\sin\!\left(\mathrm{\alpha_2}\right)}^4 + 8\, k\, L^2\, \cos^2 \: \alpha_1\, \sin^2 \: \alpha_2 +\\
    & \quad  6\, k\, L^2\, \cos^2 \: \alpha_1 + 4\, k\, L^2\, \cos\!\left(\mathrm{\alpha_1}\right)\, \cos^4 \: \alpha_2 + 2\, k\, L^2\, \cos\!\left(\mathrm{\alpha_1}\right)\, \cos^3 \: \alpha_2 +  \\
    & \quad 4\, k\, L^2\, \cos\!\left(\mathrm{\alpha_1}\right)\, \cos^2 \: \alpha_2\, \sin^2 \: \alpha_1 - 2\, k\, L^2\, \cos\!\left(\mathrm{\alpha_1}\right)\, \cos^2 \: \alpha_2\, \sin\!\left(\mathrm{\alpha_1}\right)\, \sin\!\left(\mathrm{\alpha_2}\right) +  \\
    & \quad 8\, k\, L^2\, \cos\!\left(\mathrm{\alpha_1}\right)\, \cos^2 \: \alpha_2\, \sin^2 \: \alpha_2 + 6\, k\, L^2\, \cos\!\left(\mathrm{\alpha_1}\right)\, \cos^2 \: \alpha_2 + 2\, k\, L^2\, \cos\!\left(\mathrm{\alpha_1}\right)\, \cos\!\left(\mathrm{\alpha_2}\right)\, \sin^2 \: \alpha_1 -  \\
    & \quad 2\, k\, L^2\, \cos\!\left(\mathrm{\alpha_1}\right)\, \cos\!\left(\mathrm{\alpha_2}\right)\, \sin\!\left(\mathrm{\alpha_1}\right)\, \sin\!\left(\mathrm{\alpha_2}\right) + 2\, k\, L^2\, \cos\!\left(\mathrm{\alpha_1}\right)\, \cos\!\left(\mathrm{\alpha_2}\right)\, \sin^2 \: \alpha_2 +  \\
    & \quad 2\, k\, L^2\, \cos\!\left(\mathrm{\alpha_1}\right)\, \cos\!\left(\mathrm{\alpha_2}\right) + 2\, k\, L^2\, \cos\!\left(\mathrm{\alpha_1}\right)\, \sin^2 \: \alpha_1\, \sin^2 \: \alpha_2 + 2\, k\, L^2\, \cos\!\left(\mathrm{\alpha_1}\right)\, \sin^2 \: \alpha_1 -  \\
    & \quad 2\, k\, L^2\, \cos\!\left(\mathrm{\alpha_1}\right)\, \sin\!\left(\mathrm{\alpha_1}\right)\, {\sin\!\left(\mathrm{\alpha_2}\right)}^3 - 2\, k\, L^2\, \cos\!\left(\mathrm{\alpha_1}\right)\, \sin\!\left(\mathrm{\alpha_1}\right)\, \sin\!\left(\mathrm{\alpha_2}\right) + 4\, k\, L^2\, \cos\!\left(\mathrm{\alpha_1}\right)\, {\sin\!\left(\mathrm{\alpha_2}\right)}^4 +  \\
    & \quad 6\, k\, L^2\, \cos\!\left(\mathrm{\alpha_1}\right)\, \sin^2 \: \alpha_2 + 2\, k\, L^2\, \cos\!\left(\mathrm{\alpha_1}\right) + 2\, k\, L^2\, \cos^4 \: \alpha_2\, \sin^2 \: \alpha_1 + 3\, k\, L^2\, \cos^4 \: \alpha_2 +  \\
    & \quad 2\, k\, L^2\, \cos^3 \: \alpha_2\, \sin^2 \: \alpha_1 + 2\, k\, L^2\, \cos^3 \: \alpha_2 + 4\, k\, L^2\, \cos^2 \: \alpha_2\, \sin^4 \: \alpha_1 +  \\
    & \quad 4\, k\, L^2\, \cos^2 \: \alpha_2\, \sin^2 \: \alpha_1\, \sin^2 \: \alpha_2 + 8\, k\, L^2\, \cos^2 \: \alpha_2\, \sin^2 \: \alpha_1 - 2\, k\, L^2\, \cos^2 \: \alpha_2\, \sin\!\left(\mathrm{\alpha_1}\right)\, \sin\!\left(\mathrm{\alpha_2}\right) +  \\
    & \quad 6\, k\, L^2\, \cos^2 \: \alpha_2\, \sin^2 \: \alpha_2 + 6\, k\, L^2\, \cos^2 \: \alpha_2 + 4\, k\, L^2\, \cos\!\left(\mathrm{\alpha_2}\right)\, \sin^4 \: \alpha_1 - 2\, k\, L^2\, \cos\!\left(\mathrm{\alpha_2}\right)\, \sin^3 \: \alpha_1\, \sin\!\left(\mathrm{\alpha_2}\right) + \\
    & \quad  2\, k\, L^2\, \cos\!\left(\mathrm{\alpha_2}\right)\, \sin^2 \: \alpha_1\, \sin^2 \: \alpha_2 + 6\, k\, L^2\, \cos\!\left(\mathrm{\alpha_2}\right)\, \sin^2 \: \alpha_1 - 2\, k\, L^2\, \cos\!\left(\mathrm{\alpha_2}\right)\, \sin\!\left(\mathrm{\alpha_1}\right)\, \sin\!\left(\mathrm{\alpha_2}\right) +  \\
    & \quad 2\, k\, L^2\, \cos\!\left(\mathrm{\alpha_2}\right)\, \sin^2 \: \alpha_2 + 2\, k\, L^2\, \cos\!\left(\mathrm{\alpha_2}\right) + 2\, k\, L^2\, \sin^4 \: \alpha_1\, \sin^2 \: \alpha_2 + 3\, k\, L^2\, \sin^4 \: \alpha_1 -  \\
    & \quad 2\, k\, L^2\, \sin^3 \: \alpha_1\, \sin\!\left(\mathrm{\alpha_2}\right) + 2\, k\, L^2\, \sin^2 \: \alpha_1\, {\sin\!\left(\mathrm{\alpha_2}\right)}^4 + 7\, k\, L^2\, \sin^2 \: \alpha_1\, \sin^2 \: \alpha_2 + \\
    & \quad  6\, k\, L^2\, \sin^2 \: \alpha_1 - 2\, k\, L^2\, \sin\!\left(\mathrm{\alpha_1}\right)\, {\sin\!\left(\mathrm{\alpha_2}\right)}^3 - 4\, k\, L^2\, \sin\!\left(\mathrm{\alpha_1}\right)\, \sin\!\left(\mathrm{\alpha_2}\right) + 3\, k\, L^2\, {\sin\!\left(\mathrm{\alpha_2}\right)}^4 +  \\
    & \quad 6\, k\, L^2\, \sin^2 \: \alpha_2 + 3\, k\, L^2
\end{align*}
\begin{align*}
    N_3^{31} &= - L^2\, k\, \cos\!\left(\mathrm{\alpha_1}\right)\, \left(2\, \cos^4 \: \alpha_1 + 2\, \cos^3 \: \alpha_1 - 2\, \cos^2 \: \alpha_1\, \cos\!\left(\mathrm{\alpha_2}\right) + 4\, \cos^2 \: \alpha_1\, \sin^2 \: \alpha_1 + \right.\\
    & \left. \quad 3\, \cos^2 \: \alpha_1\, \sin^2 \: \alpha_2 + 2\, \cos^2 \: \alpha_1 + 2\, \cos\!\left(\mathrm{\alpha_1}\right)\, \cos^2 \: \alpha_2 + 2\, \cos\!\left(\mathrm{\alpha_1}\right)\, \sin^2 \: \alpha_1 - \right.\\
    & \left. \quad 2\, \cos\!\left(\mathrm{\alpha_1}\right)\, \sin\!\left(\mathrm{\alpha_1}\right)\, \sin\!\left(\mathrm{\alpha_2}\right) + 4\, \cos\!\left(\mathrm{\alpha_1}\right)\, \sin^2 \: \alpha_2 + 2\, \cos\!\left(\mathrm{\alpha_1}\right) - 2\, \cos^4 \: \alpha_2 - 2\, \cos^3 \: \alpha_2 - \right.\\
    & \left. \quad 3\, \cos^2 \: \alpha_2\, \sin^2 \: \alpha_1 - 4\, \cos^2 \: \alpha_2\, \sin^2 \: \alpha_2 - 2\, \cos^2 \: \alpha_2 - 4\, \cos\!\left(\mathrm{\alpha_2}\right)\, \sin^2 \: \alpha_1 + \right.\\
    & \left. \quad 2\, \cos\!\left(\mathrm{\alpha_2}\right)\, \sin\!\left(\mathrm{\alpha_1}\right)\, \sin\!\left(\mathrm{\alpha_2}\right) - 2\, \cos\!\left(\mathrm{\alpha_2}\right)\, \sin^2 \: \alpha_2 - \right.\\
    & \left. \quad 2\, \cos\!\left(\mathrm{\alpha_2}\right) + 2\, \sin^4 \: \alpha_1 + \sin^2 \: \alpha_1 - 2\, {\sin\!\left(\mathrm{\alpha_2}\right)}^4 - \sin^2 \: \alpha_2\right)
\end{align*}
\begin{align*}
    D_3^{31} &= 8\, k\, L^2\, \cos^4 \: \alpha_1\, \cos^2 \: \alpha_2 + 8\, k\, L^2\, \cos^4 \: \alpha_1\, \cos\!\left(\mathrm{\alpha_2}\right) + 4\, k\, L^2\, \cos^4 \: \alpha_1\, \sin^2 \: \alpha_2 + 6\, k\, L^2\, \cos^4 \: \alpha_1 +  \\
    & \quad 8\, k\, L^2\, \cos^3 \: \alpha_1\, \cos^2 \: \alpha_2 + 4\, k\, L^2\, \cos^3 \: \alpha_1\, \cos\!\left(\mathrm{\alpha_2}\right) + 4\, k\, L^2\, \cos^3 \: \alpha_1\, \sin^2 \: \alpha_2 +  \\
    & \quad 4\, k\, L^2\, \cos^3 \: \alpha_1 + 8\, k\, L^2\, \cos^2 \: \alpha_1\, \cos^4 \: \alpha_2 + 8\, k\, L^2\, \cos^2 \: \alpha_1\, \cos^3 \: \alpha_2 +  \\
    & \quad 16\, k\, L^2\, \cos^2 \: \alpha_1\, \cos^2 \: \alpha_2\, \sin^2 \: \alpha_1 + 16\, k\, L^2\, \cos^2 \: \alpha_1\, \cos^2 \: \alpha_2\, \sin^2 \: \alpha_2 +  \\
    & \quad 20\, k\, L^2\, \cos^2 \: \alpha_1\, \cos^2 \: \alpha_2 + 16\, k\, L^2\, \cos^2 \: \alpha_1\, \cos\!\left(\mathrm{\alpha_2}\right)\, \sin^2 \: \alpha_1 -  \\
    & \quad 4\, k\, L^2\, \cos^2 \: \alpha_1\, \cos\!\left(\mathrm{\alpha_2}\right)\, \sin\!\left(\mathrm{\alpha_1}\right)\, \sin\!\left(\mathrm{\alpha_2}\right) + 8\, k\, L^2\, \cos^2 \: \alpha_1\, \cos\!\left(\mathrm{\alpha_2}\right)\, \sin^2 \: \alpha_2 +  \\
    & \quad 12\, k\, L^2\, \cos^2 \: \alpha_1\, \cos\!\left(\mathrm{\alpha_2}\right) + 8\, k\, L^2\, \cos^2 \: \alpha_1\, \sin^2 \: \alpha_1\, \sin^2 \: \alpha_2 + 12\, k\, L^2\, \cos^2 \: \alpha_1\, \sin^2 \: \alpha_1 -  \\
    & \quad 4\, k\, L^2\, \cos^2 \: \alpha_1\, \sin\!\left(\mathrm{\alpha_1}\right)\, \sin\!\left(\mathrm{\alpha_2}\right) + 8\, k\, L^2\, \cos^2 \: \alpha_1\, {\sin\!\left(\mathrm{\alpha_2}\right)}^4 + 16\, k\, L^2\, \cos^2 \: \alpha_1\, \sin^2 \: \alpha_2 +  \\
    & \quad 12\, k\, L^2\, \cos^2 \: \alpha_1 + 8\, k\, L^2\, \cos\!\left(\mathrm{\alpha_1}\right)\, \cos^4 \: \alpha_2 + 4\, k\, L^2\, \cos\!\left(\mathrm{\alpha_1}\right)\, \cos^3 \: \alpha_2 +  \\
    & \quad 8\, k\, L^2\, \cos\!\left(\mathrm{\alpha_1}\right)\, \cos^2 \: \alpha_2\, \sin^2 \: \alpha_1 - 4\, k\, L^2\, \cos\!\left(\mathrm{\alpha_1}\right)\, \cos^2 \: \alpha_2\, \sin\!\left(\mathrm{\alpha_1}\right)\, \sin\!\left(\mathrm{\alpha_2}\right) +  \\
    & \quad 16\, k\, L^2\, \cos\!\left(\mathrm{\alpha_1}\right)\, \cos^2 \: \alpha_2\, \sin^2 \: \alpha_2 + 12\, k\, L^2\, \cos\!\left(\mathrm{\alpha_1}\right)\, \cos^2 \: \alpha_2 + 4\, k\, L^2\, \cos\!\left(\mathrm{\alpha_1}\right)\, \cos\!\left(\mathrm{\alpha_2}\right)\, \sin^2 \: \alpha_1 -  \\
    & \quad 4\, k\, L^2\, \cos\!\left(\mathrm{\alpha_1}\right)\, \cos\!\left(\mathrm{\alpha_2}\right)\, \sin\!\left(\mathrm{\alpha_1}\right)\, \sin\!\left(\mathrm{\alpha_2}\right) + 4\, k\, L^2\, \cos\!\left(\mathrm{\alpha_1}\right)\, \cos\!\left(\mathrm{\alpha_2}\right)\, \sin^2 \: \alpha_2 +  \\
    & \quad 4\, k\, L^2\, \cos\!\left(\mathrm{\alpha_1}\right)\, \cos\!\left(\mathrm{\alpha_2}\right) + 4\, k\, L^2\, \cos\!\left(\mathrm{\alpha_1}\right)\, \sin^2 \: \alpha_1\, \sin^2 \: \alpha_2 + 4\, k\, L^2\, \cos\!\left(\mathrm{\alpha_1}\right)\, \sin^2 \: \alpha_1 -  \\
    & \quad 4\, k\, L^2\, \cos\!\left(\mathrm{\alpha_1}\right)\, \sin\!\left(\mathrm{\alpha_1}\right)\, {\sin\!\left(\mathrm{\alpha_2}\right)}^3 - 4\, k\, L^2\, \cos\!\left(\mathrm{\alpha_1}\right)\, \sin\!\left(\mathrm{\alpha_1}\right)\, \sin\!\left(\mathrm{\alpha_2}\right) + 8\, k\, L^2\, \cos\!\left(\mathrm{\alpha_1}\right)\, {\sin\!\left(\mathrm{\alpha_2}\right)}^4 +  \\
    & \quad 12\, k\, L^2\, \cos\!\left(\mathrm{\alpha_1}\right)\, \sin^2 \: \alpha_2 + 4\, k\, L^2\, \cos\!\left(\mathrm{\alpha_1}\right) + 4\, k\, L^2\, \cos^4 \: \alpha_2\, \sin^2 \: \alpha_1 + 6\, k\, L^2\, \cos^4 \: \alpha_2 +  \\
    & \quad 4\, k\, L^2\, \cos^3 \: \alpha_2\, \sin^2 \: \alpha_1 + 4\, k\, L^2\, \cos^3 \: \alpha_2 + 8\, k\, L^2\, \cos^2 \: \alpha_2\, \sin^4 \: \alpha_1 +  \\
    & \quad 8\, k\, L^2\, \cos^2 \: \alpha_2\, \sin^2 \: \alpha_1\, \sin^2 \: \alpha_2 + 16\, k\, L^2\, \cos^2 \: \alpha_2\, \sin^2 \: \alpha_1 - 4\, k\, L^2\, \cos^2 \: \alpha_2\, \sin\!\left(\mathrm{\alpha_1}\right)\, \sin\!\left(\mathrm{\alpha_2}\right) +  \\
    & \quad 12\, k\, L^2\, \cos^2 \: \alpha_2\, \sin^2 \: \alpha_2 + 12\, k\, L^2\, \cos^2 \: \alpha_2 + 8\, k\, L^2\, \cos\!\left(\mathrm{\alpha_2}\right)\, \sin^4 \: \alpha_1 -  \\
    & \quad 4\, k\, L^2\, \cos\!\left(\mathrm{\alpha_2}\right)\, \sin^3 \: \alpha_1\, \sin\!\left(\mathrm{\alpha_2}\right) + 4\, k\, L^2\, \cos\!\left(\mathrm{\alpha_2}\right)\, \sin^2 \: \alpha_1\, \sin^2 \: \alpha_2 + 12\, k\, L^2\, \cos\!\left(\mathrm{\alpha_2}\right)\, \sin^2 \: \alpha_1 -  \\
    & \quad 4\, k\, L^2\, \cos\!\left(\mathrm{\alpha_2}\right)\, \sin\!\left(\mathrm{\alpha_1}\right)\, \sin\!\left(\mathrm{\alpha_2}\right) + 4\, k\, L^2\, \cos\!\left(\mathrm{\alpha_2}\right)\, \sin^2 \: \alpha_2 + 4\, k\, L^2\, \cos\!\left(\mathrm{\alpha_2}\right) +  \\
    & \quad 4\, k\, L^2\, \sin^4 \: \alpha_1\, \sin^2 \: \alpha_2 + 6\, k\, L^2\, \sin^4 \: \alpha_1 - 4\, k\, L^2\, \sin^3 \: \alpha_1\, \sin\!\left(\mathrm{\alpha_2}\right) + 4\, k\, L^2\, \sin^2 \: \alpha_1\, {\sin\!\left(\mathrm{\alpha_2}\right)}^4 +  \\
    & \quad 14\, k\, L^2\, \sin^2 \: \alpha_1\, \sin^2 \: \alpha_2 + 12\, k\, L^2\, \sin^2 \: \alpha_1 - 4\, k\, L^2\, \sin\!\left(\mathrm{\alpha_1}\right)\, {\sin\!\left(\mathrm{\alpha_2}\right)}^3 - 8\, k\, L^2\, \sin\!\left(\mathrm{\alpha_1}\right)\, \sin\!\left(\mathrm{\alpha_2}\right) +  \\
    & \quad 6\, k\, L^2\, {\sin\!\left(\mathrm{\alpha_2}\right)}^4 + 12\, k\, L^2\, \sin^2 \: \alpha_2 + 6\, k\, L^2
\end{align*}

\begin{align*}
    N_1^{32} &= L^2\, k\, \sin\!\left(\mathrm{\alpha_2}\right)\, \left(2\, \cos^2 \: \alpha_1\, \cos\!\left(\mathrm{\alpha_2}\right)\, \sin\!\left(\mathrm{\alpha_2}\right) + \cos^2 \: \alpha_1\, \sin\!\left(\mathrm{\alpha_1}\right) + 2\, \cos^2 \: \alpha_1\, \sin\!\left(\mathrm{\alpha_2}\right) + \right.\\
    & \left. \quad 2\, \cos\!\left(\mathrm{\alpha_1}\right)\, \cos^2 \: \alpha_2\, \sin\!\left(\mathrm{\alpha_1}\right) + \cos\!\left(\mathrm{\alpha_1}\right)\, \cos\!\left(\mathrm{\alpha_2}\right)\, \sin\!\left(\mathrm{\alpha_1}\right) + \cos\!\left(\mathrm{\alpha_1}\right)\, \cos\!\left(\mathrm{\alpha_2}\right)\, \sin\!\left(\mathrm{\alpha_2}\right) +\right.\\
    & \left. \quad  \cos\!\left(\mathrm{\alpha_1}\right)\, \sin\!\left(\mathrm{\alpha_1}\right)\, \sin^2 \: \alpha_2 + \cos\!\left(\mathrm{\alpha_1}\right)\, \sin\!\left(\mathrm{\alpha_1}\right) + 2\, \cos^2 \: \alpha_2\, \sin\!\left(\mathrm{\alpha_1}\right) + \right.\\
    & \left. \quad \cos^2 \: \alpha_2\, \sin\!\left(\mathrm{\alpha_2}\right) + \cos\!\left(\mathrm{\alpha_2}\right)\, \sin^2 \: \alpha_1\, \sin\!\left(\mathrm{\alpha_2}\right) + \cos\!\left(\mathrm{\alpha_2}\right)\, \sin\!\left(\mathrm{\alpha_2}\right) + \sin^3 \: \alpha_1 + \sin^2 \: \alpha_1\, \sin\!\left(\mathrm{\alpha_2}\right) + \right.\\
    & \left. \quad \sin\!\left(\mathrm{\alpha_1}\right)\, \sin^2 \: \alpha_2 + 2\, \sin\!\left(\mathrm{\alpha_1}\right) + {\sin\!\left(\mathrm{\alpha_2}\right)}^3 + 2\, \sin\!\left(\mathrm{\alpha_2}\right)\right)
\end{align*}
\begin{align*}
    D_1^{32} &= 4\, k\, L^2\, \cos^4 \: \alpha_1\, \cos^2 \: \alpha_2 + 4\, k\, L^2\, \cos^4 \: \alpha_1\, \cos\!\left(\mathrm{\alpha_2}\right) + 2\, k\, L^2\, \cos^4 \: \alpha_1\, \sin^2 \: \alpha_2 + 3\, k\, L^2\, \cos^4 \: \alpha_1 +  \\
    & \quad 4\, k\, L^2\, \cos^3 \: \alpha_1\, \cos^2 \: \alpha_2 + 2\, k\, L^2\, \cos^3 \: \alpha_1\, \cos\!\left(\mathrm{\alpha_2}\right) + 2\, k\, L^2\, \cos^3 \: \alpha_1\, \sin^2 \: \alpha_2 + 2\, k\, L^2\, \cos^3 \: \alpha_1 +  \\
    & \quad 4\, k\, L^2\, \cos^2 \: \alpha_1\, \cos^4 \: \alpha_2 + 4\, k\, L^2\, \cos^2 \: \alpha_1\, \cos^3 \: \alpha_2 + 8\, k\, L^2\, \cos^2 \: \alpha_1\, \cos^2 \: \alpha_2\, \sin^2 \: \alpha_1 +  \\
    & \quad 8\, k\, L^2\, \cos^2 \: \alpha_1\, \cos^2 \: \alpha_2\, \sin^2 \: \alpha_2 + 10\, k\, L^2\, \cos^2 \: \alpha_1\, \cos^2 \: \alpha_2 + 8\, k\, L^2\, \cos^2 \: \alpha_1\, \cos\!\left(\mathrm{\alpha_2}\right)\, \sin^2 \: \alpha_1 -  \\
    & \quad 2\, k\, L^2\, \cos^2 \: \alpha_1\, \cos\!\left(\mathrm{\alpha_2}\right)\, \sin\!\left(\mathrm{\alpha_1}\right)\, \sin\!\left(\mathrm{\alpha_2}\right) + 4\, k\, L^2\, \cos^2 \: \alpha_1\, \cos\!\left(\mathrm{\alpha_2}\right)\, \sin^2 \: \alpha_2 +  \\
    & \quad 6\, k\, L^2\, \cos^2 \: \alpha_1\, \cos\!\left(\mathrm{\alpha_2}\right) + 4\, k\, L^2\, \cos^2 \: \alpha_1\, \sin^2 \: \alpha_1\, \sin^2 \: \alpha_2 + 6\, k\, L^2\, \cos^2 \: \alpha_1\, \sin^2 \: \alpha_1 -  \\
    & \quad 2\, k\, L^2\, \cos^2 \: \alpha_1\, \sin\!\left(\mathrm{\alpha_1}\right)\, \sin\!\left(\mathrm{\alpha_2}\right) + 4\, k\, L^2\, \cos^2 \: \alpha_1\, {\sin\!\left(\mathrm{\alpha_2}\right)}^4 + 8\, k\, L^2\, \cos^2 \: \alpha_1\, \sin^2 \: \alpha_2 + \\
    & \quad  6\, k\, L^2\, \cos^2 \: \alpha_1 + 4\, k\, L^2\, \cos\!\left(\mathrm{\alpha_1}\right)\, \cos^4 \: \alpha_2 + 2\, k\, L^2\, \cos\!\left(\mathrm{\alpha_1}\right)\, \cos^3 \: \alpha_2 +  \\
    & \quad 4\, k\, L^2\, \cos\!\left(\mathrm{\alpha_1}\right)\, \cos^2 \: \alpha_2\, \sin^2 \: \alpha_1 - 2\, k\, L^2\, \cos\!\left(\mathrm{\alpha_1}\right)\, \cos^2 \: \alpha_2\, \sin\!\left(\mathrm{\alpha_1}\right)\, \sin\!\left(\mathrm{\alpha_2}\right) +  \\
    & \quad 8\, k\, L^2\, \cos\!\left(\mathrm{\alpha_1}\right)\, \cos^2 \: \alpha_2\, \sin^2 \: \alpha_2 + 6\, k\, L^2\, \cos\!\left(\mathrm{\alpha_1}\right)\, \cos^2 \: \alpha_2 + 2\, k\, L^2\, \cos\!\left(\mathrm{\alpha_1}\right)\, \cos\!\left(\mathrm{\alpha_2}\right)\, \sin^2 \: \alpha_1 -  \\
    & \quad 2\, k\, L^2\, \cos\!\left(\mathrm{\alpha_1}\right)\, \cos\!\left(\mathrm{\alpha_2}\right)\, \sin\!\left(\mathrm{\alpha_1}\right)\, \sin\!\left(\mathrm{\alpha_2}\right) + 2\, k\, L^2\, \cos\!\left(\mathrm{\alpha_1}\right)\, \cos\!\left(\mathrm{\alpha_2}\right)\, \sin^2 \: \alpha_2 +  \\
    & \quad 2\, k\, L^2\, \cos\!\left(\mathrm{\alpha_1}\right)\, \cos\!\left(\mathrm{\alpha_2}\right) + 2\, k\, L^2\, \cos\!\left(\mathrm{\alpha_1}\right)\, \sin^2 \: \alpha_1\, \sin^2 \: \alpha_2 + 2\, k\, L^2\, \cos\!\left(\mathrm{\alpha_1}\right)\, \sin^2 \: \alpha_1 -  \\
    & \quad 2\, k\, L^2\, \cos\!\left(\mathrm{\alpha_1}\right)\, \sin\!\left(\mathrm{\alpha_1}\right)\, {\sin\!\left(\mathrm{\alpha_2}\right)}^3 - 2\, k\, L^2\, \cos\!\left(\mathrm{\alpha_1}\right)\, \sin\!\left(\mathrm{\alpha_1}\right)\, \sin\!\left(\mathrm{\alpha_2}\right) + 4\, k\, L^2\, \cos\!\left(\mathrm{\alpha_1}\right)\, {\sin\!\left(\mathrm{\alpha_2}\right)}^4 +  \\
    & \quad 6\, k\, L^2\, \cos\!\left(\mathrm{\alpha_1}\right)\, \sin^2 \: \alpha_2 + 2\, k\, L^2\, \cos\!\left(\mathrm{\alpha_1}\right) + 2\, k\, L^2\, \cos^4 \: \alpha_2\, \sin^2 \: \alpha_1 + 3\, k\, L^2\, \cos^4 \: \alpha_2 +  \\
    & \quad 2\, k\, L^2\, \cos^3 \: \alpha_2\, \sin^2 \: \alpha_1 + 2\, k\, L^2\, \cos^3 \: \alpha_2 + 4\, k\, L^2\, \cos^2 \: \alpha_2\, \sin^4 \: \alpha_1 +  \\
    & \quad 4\, k\, L^2\, \cos^2 \: \alpha_2\, \sin^2 \: \alpha_1\, \sin^2 \: \alpha_2 + 8\, k\, L^2\, \cos^2 \: \alpha_2\, \sin^2 \: \alpha_1 - 2\, k\, L^2\, \cos^2 \: \alpha_2\, \sin\!\left(\mathrm{\alpha_1}\right)\, \sin\!\left(\mathrm{\alpha_2}\right) +  \\
    & \quad 6\, k\, L^2\, \cos^2 \: \alpha_2\, \sin^2 \: \alpha_2 + 6\, k\, L^2\, \cos^2 \: \alpha_2 + 4\, k\, L^2\, \cos\!\left(\mathrm{\alpha_2}\right)\, \sin^4 \: \alpha_1 -  \\
    & \quad 2\, k\, L^2\, \cos\!\left(\mathrm{\alpha_2}\right)\, \sin^3 \: \alpha_1\, \sin\!\left(\mathrm{\alpha_2}\right) + 2\, k\, L^2\, \cos\!\left(\mathrm{\alpha_2}\right)\, \sin^2 \: \alpha_1\, \sin^2 \: \alpha_2 + 6\, k\, L^2\, \cos\!\left(\mathrm{\alpha_2}\right)\, \sin^2 \: \alpha_1 -  \\
    & \quad 2\, k\, L^2\, \cos\!\left(\mathrm{\alpha_2}\right)\, \sin\!\left(\mathrm{\alpha_1}\right)\, \sin\!\left(\mathrm{\alpha_2}\right) + 2\, k\, L^2\, \cos\!\left(\mathrm{\alpha_2}\right)\, \sin^2 \: \alpha_2 + 2\, k\, L^2\, \cos\!\left(\mathrm{\alpha_2}\right) +  \\
    & \quad 2\, k\, L^2\, \sin^4 \: \alpha_1\, \sin^2 \: \alpha_2 + 3\, k\, L^2\, \sin^4 \: \alpha_1 - 2\, k\, L^2\, \sin^3 \: \alpha_1\, \sin\!\left(\mathrm{\alpha_2}\right) + 2\, k\, L^2\, \sin^2 \: \alpha_1\, {\sin\!\left(\mathrm{\alpha_2}\right)}^4 +  \\
    & \quad 7\, k\, L^2\, \sin^2 \: \alpha_1\, \sin^2 \: \alpha_2 + 6\, k\, L^2\, \sin^2 \: \alpha_1 - 2\, k\, L^2\, \sin\!\left(\mathrm{\alpha_1}\right)\, {\sin\!\left(\mathrm{\alpha_2}\right)}^3 - 4\, k\, L^2\, \sin\!\left(\mathrm{\alpha_1}\right)\, \sin\!\left(\mathrm{\alpha_2}\right) +  \\
    & \quad 3\, k\, L^2\, {\sin\!\left(\mathrm{\alpha_2}\right)}^4 + 6\, k\, L^2\, \sin^2 \: \alpha_2 + 3\, k\, L^2
\end{align*}

\begin{align*}
    N_2^{32} &= - \left(\frac{2\, L^3\, k}{3} + 2\, L^3\, k\, \left(\cos\!\left(\mathrm{\alpha_2}\right) + 1\right)\right)\, \left(2\, \cos^4 \: \alpha_1 + 4\, \cos^2 \: \alpha_1\, \cos^2 \: \alpha_2 + 4\, \cos^2 \: \alpha_1\, \sin^2 \: \alpha_1 + \right.\\
    & \left. \quad 5\, \cos^2 \: \alpha_1\, \sin^2 \: \alpha_2 + 4\, \cos^2 \: \alpha_1 + 2\, \cos\!\left(\mathrm{\alpha_1}\right)\, \cos\!\left(\mathrm{\alpha_2}\right)\, \sin\!\left(\mathrm{\alpha_1}\right)\, \sin\!\left(\mathrm{\alpha_2}\right) + 2\, \cos^4 \: \alpha_2 + \right.\\
    & \left. \quad 5\, \cos^2 \: \alpha_2\, \sin^2 \: \alpha_1 + 4\, \cos^2 \: \alpha_2\, \sin^2 \: \alpha_2 + 4\, \cos^2 \: \alpha_2 + 2\, \sin^4 \: \alpha_1 + \right.\\
    & \left. \quad 4\, \sin^2 \: \alpha_1\, \sin^2 \: \alpha_2 + 5\, \sin^2 \: \alpha_1 + 2\, {\sin\!\left(\mathrm{\alpha_2}\right)}^4 + 5\, \sin^2 \: \alpha_2 + 2\right)
\end{align*}
\begin{align*}
    D_2^{32} &= 16\, k\, L^3\, \cos^4 \: \alpha_1\, \cos^2 \: \alpha_2 + 16\, k\, L^3\, \cos^4 \: \alpha_1\, \cos\!\left(\mathrm{\alpha_2}\right) + 8\, k\, L^3\, \cos^4 \: \alpha_1\, \sin^2 \: \alpha_2 + 12\, k\, L^3\, \cos^4 \: \alpha_1 + \\
    & \quad  16\, k\, L^3\, \cos^3 \: \alpha_1\, \cos^2 \: \alpha_2 + 8\, k\, L^3\, \cos^3 \: \alpha_1\, \cos\!\left(\mathrm{\alpha_2}\right) + 8\, k\, L^3\, \cos^3 \: \alpha_1\, \sin^2 \: \alpha_2 + 8\, k\, L^3\, \cos^3 \: \alpha_1 + \\
    & \quad 16\, k\, L^3\, \cos^2 \: \alpha_1\, \cos^4 \: \alpha_2 + 16\, k\, L^3\, \cos^2 \: \alpha_1\, \cos^3 \: \alpha_2 + 32\, k\, L^3\, \cos^2 \: \alpha_1\, \cos^2 \: \alpha_2\, \sin^2 \: \alpha_1 + \\
    & \quad 32\, k\, L^3\, \cos^2 \: \alpha_1\, \cos^2 \: \alpha_2\, \sin^2 \: \alpha_2 + 40\, k\, L^3\, \cos^2 \: \alpha_1\, \cos^2 \: \alpha_2 + 32\, k\, L^3\, \cos^2 \: \alpha_1\, \cos\!\left(\mathrm{\alpha_2}\right)\, \sin^2 \: \alpha_1 - \\
    & \quad 8\, k\, L^3\, \cos^2 \: \alpha_1\, \cos\!\left(\mathrm{\alpha_2}\right)\, \sin\!\left(\mathrm{\alpha_1}\right)\, \sin\!\left(\mathrm{\alpha_2}\right) + 16\, k\, L^3\, \cos^2 \: \alpha_1\, \cos\!\left(\mathrm{\alpha_2}\right)\, \sin^2 \: \alpha_2 + \\
    & \quad 24\, k\, L^3\, \cos^2 \: \alpha_1\, \cos\!\left(\mathrm{\alpha_2}\right) + 16\, k\, L^3\, \cos^2 \: \alpha_1\, \sin^2 \: \alpha_1\, \sin^2 \: \alpha_2 + 24\, k\, L^3\, \cos^2 \: \alpha_1\, \sin^2 \: \alpha_1 - \\
    & \quad 8\, k\, L^3\, \cos^2 \: \alpha_1\, \sin\!\left(\mathrm{\alpha_1}\right)\, \sin\!\left(\mathrm{\alpha_2}\right) + 16\, k\, L^3\, \cos^2 \: \alpha_1\, {\sin\!\left(\mathrm{\alpha_2}\right)}^4 + 32\, k\, L^3\, \cos^2 \: \alpha_1\, \sin^2 \: \alpha_2 + \\
    & \quad 24\, k\, L^3\, \cos^2 \: \alpha_1 + 16\, k\, L^3\, \cos\!\left(\mathrm{\alpha_1}\right)\, \cos^4 \: \alpha_2 + 8\, k\, L^3\, \cos\!\left(\mathrm{\alpha_1}\right)\, \cos^3 \: \alpha_2 + \\
    & \quad 16\, k\, L^3\, \cos\!\left(\mathrm{\alpha_1}\right)\, \cos^2 \: \alpha_2\, \sin^2 \: \alpha_1 - 8\, k\, L^3\, \cos\!\left(\mathrm{\alpha_1}\right)\, \cos^2 \: \alpha_2\, \sin\!\left(\mathrm{\alpha_1}\right)\, \sin\!\left(\mathrm{\alpha_2}\right) + \\
    & \quad 32\, k\, L^3\, \cos\!\left(\mathrm{\alpha_1}\right)\, \cos^2 \: \alpha_2\, \sin^2 \: \alpha_2 + 24\, k\, L^3\, \cos\!\left(\mathrm{\alpha_1}\right)\, \cos^2 \: \alpha_2 + 8\, k\, L^3\, \cos\!\left(\mathrm{\alpha_1}\right)\, \cos\!\left(\mathrm{\alpha_2}\right)\, \sin^2 \: \alpha_1 - \\
    & \quad 8\, k\, L^3\, \cos\!\left(\mathrm{\alpha_1}\right)\, \cos\!\left(\mathrm{\alpha_2}\right)\, \sin\!\left(\mathrm{\alpha_1}\right)\, \sin\!\left(\mathrm{\alpha_2}\right) + 8\, k\, L^3\, \cos\!\left(\mathrm{\alpha_1}\right)\, \cos\!\left(\mathrm{\alpha_2}\right)\, \sin^2 \: \alpha_2 + 8\, k\, L^3\, \cos\!\left(\mathrm{\alpha_1}\right)\, \cos\!\left(\mathrm{\alpha_2}\right) + \\
    & \quad 8\, k\, L^3\, \cos\!\left(\mathrm{\alpha_1}\right)\, \sin^2 \: \alpha_1\, \sin^2 \: \alpha_2 + 8\, k\, L^3\, \cos\!\left(\mathrm{\alpha_1}\right)\, \sin^2 \: \alpha_1 - 8\, k\, L^3\, \cos\!\left(\mathrm{\alpha_1}\right)\, \sin\!\left(\mathrm{\alpha_1}\right)\, {\sin\!\left(\mathrm{\alpha_2}\right)}^3 - \\
    & \quad 8\, k\, L^3\, \cos\!\left(\mathrm{\alpha_1}\right)\, \sin\!\left(\mathrm{\alpha_1}\right)\, \sin\!\left(\mathrm{\alpha_2}\right) + 16\, k\, L^3\, \cos\!\left(\mathrm{\alpha_1}\right)\, {\sin\!\left(\mathrm{\alpha_2}\right)}^4 + \\
    & \quad 24\, k\, L^3\, \cos\!\left(\mathrm{\alpha_1}\right)\, \sin^2 \: \alpha_2 + 8\, k\, L^3\, \cos\!\left(\mathrm{\alpha_1}\right) + 8\, k\, L^3\, \cos^4 \: \alpha_2\, \sin^2 \: \alpha_1 + 12\, k\, L^3\, \cos^4 \: \alpha_2 + \\
    & \quad 8\, k\, L^3\, \cos^3 \: \alpha_2\, \sin^2 \: \alpha_1 + 8\, k\, L^3\, \cos^3 \: \alpha_2 + 16\, k\, L^3\, \cos^2 \: \alpha_2\, \sin^4 \: \alpha_1 + \\
    & \quad 16\, k\, L^3\, \cos^2 \: \alpha_2\, \sin^2 \: \alpha_1\, \sin^2 \: \alpha_2 + 32\, k\, L^3\, \cos^2 \: \alpha_2\, \sin^2 \: \alpha_1 -\\
    & \quad  8\, k\, L^3\, \cos^2 \: \alpha_2\, \sin\!\left(\mathrm{\alpha_1}\right)\, \sin\!\left(\mathrm{\alpha_2}\right) + 24\, k\, L^3\, \cos^2 \: \alpha_2\, \sin^2 \: \alpha_2 + 24\, k\, L^3\, \cos^2 \: \alpha_2 + \\
    & \quad 16\, k\, L^3\, \cos\!\left(\mathrm{\alpha_2}\right)\, \sin^4 \: \alpha_1 - 8\, k\, L^3\, \cos\!\left(\mathrm{\alpha_2}\right)\, \sin^3 \: \alpha_1\, \sin\!\left(\mathrm{\alpha_2}\right) +\\
    & \quad  8\, k\, L^3\, \cos\!\left(\mathrm{\alpha_2}\right)\, \sin^2 \: \alpha_1\, \sin^2 \: \alpha_2 + 24\, k\, L^3\, \cos\!\left(\mathrm{\alpha_2}\right)\, \sin^2 \: \alpha_1 - 8\, k\, L^3\, \cos\!\left(\mathrm{\alpha_2}\right)\, \sin\!\left(\mathrm{\alpha_1}\right)\, \sin\!\left(\mathrm{\alpha_2}\right) + \\
    & \quad 8\, k\, L^3\, \cos\!\left(\mathrm{\alpha_2}\right)\, \sin^2 \: \alpha_2 + 8\, k\, L^3\, \cos\!\left(\mathrm{\alpha_2}\right) + 8\, k\, L^3\, \sin^4 \: \alpha_1\, \sin^2 \: \alpha_2 + 12\, k\, L^3\, \sin^4 \: \alpha_1 - \\
    & \quad 8\, k\, L^3\, \sin^3 \: \alpha_1\, \sin\!\left(\mathrm{\alpha_2}\right) + 8\, k\, L^3\, \sin^2 \: \alpha_1\, {\sin\!\left(\mathrm{\alpha_2}\right)}^4 + 28\, k\, L^3\, \sin^2 \: \alpha_1\, \sin^2 \: \alpha_2 + \\
    & \quad 24\, k\, L^3\, \sin^2 \: \alpha_1 - 8\, k\, L^3\, \sin\!\left(\mathrm{\alpha_1}\right)\, {\sin\!\left(\mathrm{\alpha_2}\right)}^3 - 16\, k\, L^3\, \sin\!\left(\mathrm{\alpha_1}\right)\, \sin\!\left(\mathrm{\alpha_2}\right) + 12\, k\, L^3\, {\sin\!\left(\mathrm{\alpha_2}\right)}^4 + \\
    & \quad 24\, k\, L^3\, \sin^2 \: \alpha_2 + 12\, k\, L^3
\end{align*}
\begin{align*}
    N_3^{32} &= - L^2\, k\, \cos\!\left(\mathrm{\alpha_2}\right)\, \left(2\, \cos^4 \: \alpha_1 + 2\, \cos^3 \: \alpha_1 - 2\, \cos^2 \: \alpha_1\, \cos\!\left(\mathrm{\alpha_2}\right) + 4\, \cos^2 \: \alpha_1\, \sin^2 \: \alpha_1 + \right.\\
    & \left. \quad 3\, \cos^2 \: \alpha_1\, \sin^2 \: \alpha_2 + 2\, \cos^2 \: \alpha_1 + 2\, \cos\!\left(\mathrm{\alpha_1}\right)\, \cos^2 \: \alpha_2 + 2\, \cos\!\left(\mathrm{\alpha_1}\right)\, \sin^2 \: \alpha_1 - \right.\\
    & \left. \quad 2\, \cos\!\left(\mathrm{\alpha_1}\right)\, \sin\!\left(\mathrm{\alpha_1}\right)\, \sin\!\left(\mathrm{\alpha_2}\right) + 4\, \cos\!\left(\mathrm{\alpha_1}\right)\, \sin^2 \: \alpha_2 + 2\, \cos\!\left(\mathrm{\alpha_1}\right) - 2\, \cos^4 \: \alpha_2 - 2\, \cos^3 \: \alpha_2 - \right.\\
    & \left. \quad 3\, \cos^2 \: \alpha_2\, \sin^2 \: \alpha_1 - 4\, \cos^2 \: \alpha_2\, \sin^2 \: \alpha_2 - 2\, \cos^2 \: \alpha_2 - 4\, \cos\!\left(\mathrm{\alpha_2}\right)\, \sin^2 \: \alpha_1 + \right.\\
    & \left. \quad 2\, \cos\!\left(\mathrm{\alpha_2}\right)\, \sin\!\left(\mathrm{\alpha_1}\right)\, \sin\!\left(\mathrm{\alpha_2}\right) - 2\, \cos\!\left(\mathrm{\alpha_2}\right)\, \sin^2 \: \alpha_2 - 2\, \cos\!\left(\mathrm{\alpha_2}\right) + 2\, \sin^4 \: \alpha_1 + \right.\\
    & \left. \quad \sin^2 \: \alpha_1 - 2\, {\sin\!\left(\mathrm{\alpha_2}\right)}^4 - \sin^2 \: \alpha_2\right)
\end{align*}

\begin{align*}
    D_3^{32} &= 8\, k\, L^2\, \cos^4 \: \alpha_1\, \cos^2 \: \alpha_2 + 8\, k\, L^2\, \cos^4 \: \alpha_1\, \cos\!\left(\mathrm{\alpha_2}\right) + 4\, k\, L^2\, \cos^4 \: \alpha_1\, \sin^2 \: \alpha_2 + \\
    & \quad 6\, k\, L^2\, \cos^4 \: \alpha_1 + 8\, k\, L^2\, \cos^3 \: \alpha_1\, \cos^2 \: \alpha_2 + 4\, k\, L^2\, \cos^3 \: \alpha_1\, \cos\!\left(\mathrm{\alpha_2}\right) +  \\
    & \quad 4\, k\, L^2\, \cos^3 \: \alpha_1\, \sin^2 \: \alpha_2 + 4\, k\, L^2\, \cos^3 \: \alpha_1 + 8\, k\, L^2\, \cos^2 \: \alpha_1\, \cos^4 \: \alpha_2 + 8\, k\, L^2\, \cos^2 \: \alpha_1\, \cos^3 \: \alpha_2 +  \\
    & \quad 16\, k\, L^2\, \cos^2 \: \alpha_1\, \cos^2 \: \alpha_2\, \sin^2 \: \alpha_1 + 16\, k\, L^2\, \cos^2 \: \alpha_1\, \cos^2 \: \alpha_2\, \sin^2 \: \alpha_2 + 20\, k\, L^2\, \cos^2 \: \alpha_1\, \cos^2 \: \alpha_2 +  \\
    & \quad 16\, k\, L^2\, \cos^2 \: \alpha_1\, \cos\!\left(\mathrm{\alpha_2}\right)\, \sin^2 \: \alpha_1 - 4\, k\, L^2\, \cos^2 \: \alpha_1\, \cos\!\left(\mathrm{\alpha_2}\right)\, \sin\!\left(\mathrm{\alpha_1}\right)\, \sin\!\left(\mathrm{\alpha_2}\right) + \\
    & \quad 8\, k\, L^2\, \cos^2 \: \alpha_1\, \cos\!\left(\mathrm{\alpha_2}\right)\, \sin^2 \: \alpha_2 + 12\, k\, L^2\, \cos^2 \: \alpha_1\, \cos\!\left(\mathrm{\alpha_2}\right) + 8\, k\, L^2\, \cos^2 \: \alpha_1\, \sin^2 \: \alpha_1\, \sin^2 \: \alpha_2 + \\
    & \quad 12\, k\, L^2\, \cos^2 \: \alpha_1\, \sin^2 \: \alpha_1 - 4\, k\, L^2\, \cos^2 \: \alpha_1\, \sin\!\left(\mathrm{\alpha_1}\right)\, \sin\!\left(\mathrm{\alpha_2}\right) + 8\, k\, L^2\, \cos^2 \: \alpha_1\, {\sin\!\left(\mathrm{\alpha_2}\right)}^4 + \\
    & \quad 16\, k\, L^2\, \cos^2 \: \alpha_1\, \sin^2 \: \alpha_2 + 12\, k\, L^2\, \cos^2 \: \alpha_1 + 8\, k\, L^2\, \cos\!\left(\mathrm{\alpha_1}\right)\, \cos^4 \: \alpha_2 + 4\, k\, L^2\, \cos\!\left(\mathrm{\alpha_1}\right)\, \cos^3 \: \alpha_2 + \\
    & \quad 8\, k\, L^2\, \cos\!\left(\mathrm{\alpha_1}\right)\, \cos^2 \: \alpha_2\, \sin^2 \: \alpha_1 - 4\, k\, L^2\, \cos\!\left(\mathrm{\alpha_1}\right)\, \cos^2 \: \alpha_2\, \sin\!\left(\mathrm{\alpha_1}\right)\, \sin\!\left(\mathrm{\alpha_2}\right) + \\
    & \quad 16\, k\, L^2\, \cos\!\left(\mathrm{\alpha_1}\right)\, \cos^2 \: \alpha_2\, \sin^2 \: \alpha_2 + 12\, k\, L^2\, \cos\!\left(\mathrm{\alpha_1}\right)\, \cos^2 \: \alpha_2 + 4\, k\, L^2\, \cos\!\left(\mathrm{\alpha_1}\right)\, \cos\!\left(\mathrm{\alpha_2}\right)\, \sin^2 \: \alpha_1 - \\
    & \quad 4\, k\, L^2\, \cos\!\left(\mathrm{\alpha_1}\right)\, \cos\!\left(\mathrm{\alpha_2}\right)\, \sin\!\left(\mathrm{\alpha_1}\right)\, \sin\!\left(\mathrm{\alpha_2}\right) + 4\, k\, L^2\, \cos\!\left(\mathrm{\alpha_1}\right)\, \cos\!\left(\mathrm{\alpha_2}\right)\, \sin^2 \: \alpha_2 + \\
    & \quad 4\, k\, L^2\, \cos\!\left(\mathrm{\alpha_1}\right)\, \cos\!\left(\mathrm{\alpha_2}\right) + 4\, k\, L^2\, \cos\!\left(\mathrm{\alpha_1}\right)\, \sin^2 \: \alpha_1\, \sin^2 \: \alpha_2 + 4\, k\, L^2\, \cos\!\left(\mathrm{\alpha_1}\right)\, \sin^2 \: \alpha_1 - \\
    & \quad 4\, k\, L^2\, \cos\!\left(\mathrm{\alpha_1}\right)\, \sin\!\left(\mathrm{\alpha_1}\right)\, {\sin\!\left(\mathrm{\alpha_2}\right)}^3 - 4\, k\, L^2\, \cos\!\left(\mathrm{\alpha_1}\right)\, \sin\!\left(\mathrm{\alpha_1}\right)\, \sin\!\left(\mathrm{\alpha_2}\right) + 8\, k\, L^2\, \cos\!\left(\mathrm{\alpha_1}\right)\, {\sin\!\left(\mathrm{\alpha_2}\right)}^4 + \\
    & \quad 12\, k\, L^2\, \cos\!\left(\mathrm{\alpha_1}\right)\, \sin^2 \: \alpha_2 + 4\, k\, L^2\, \cos\!\left(\mathrm{\alpha_1}\right) + 4\, k\, L^2\, \cos^4 \: \alpha_2\, \sin^2 \: \alpha_1 + 6\, k\, L^2\, \cos^4 \: \alpha_2 + \\
    & \quad 4\, k\, L^2\, \cos^3 \: \alpha_2\, \sin^2 \: \alpha_1 + 4\, k\, L^2\, \cos^3 \: \alpha_2 + 8\, k\, L^2\, \cos^2 \: \alpha_2\, \sin^4 \: \alpha_1 + \\
    & \quad  8\, k\, L^2\, \cos^2 \: \alpha_2\, \sin^2 \: \alpha_1\, \sin^2 \: \alpha_2 + 16\, k\, L^2\, \cos^2 \: \alpha_2\, \sin^2 \: \alpha_1 - 4\, k\, L^2\, \cos^2 \: \alpha_2\, \sin\!\left(\mathrm{\alpha_1}\right)\, \sin\!\left(\mathrm{\alpha_2}\right) + \\
    & \quad 12\, k\, L^2\, \cos^2 \: \alpha_2\, \sin^2 \: \alpha_2 + 12\, k\, L^2\, \cos^2 \: \alpha_2 + 8\, k\, L^2\, \cos\!\left(\mathrm{\alpha_2}\right)\, \sin^4 \: \alpha_1 - \\
    & \quad 4\, k\, L^2\, \cos\!\left(\mathrm{\alpha_2}\right)\, \sin^3 \: \alpha_1\, \sin\!\left(\mathrm{\alpha_2}\right) + 4\, k\, L^2\, \cos\!\left(\mathrm{\alpha_2}\right)\, \sin^2 \: \alpha_1\, \sin^2 \: \alpha_2 + 12\, k\, L^2\, \cos\!\left(\mathrm{\alpha_2}\right)\, \sin^2 \: \alpha_1 - \\
    & \quad 4\, k\, L^2\, \cos\!\left(\mathrm{\alpha_2}\right)\, \sin\!\left(\mathrm{\alpha_1}\right)\, \sin\!\left(\mathrm{\alpha_2}\right) + 4\, k\, L^2\, \cos\!\left(\mathrm{\alpha_2}\right)\, \sin^2 \: \alpha_2 + 4\, k\, L^2\, \cos\!\left(\mathrm{\alpha_2}\right) + \\
    & \quad 4\, k\, L^2\, \sin^4 \: \alpha_1\, \sin^2 \: \alpha_2 + 6\, k\, L^2\, \sin^4 \: \alpha_1 - 4\, k\, L^2\, \sin^3 \: \alpha_1\, \sin\!\left(\mathrm{\alpha_2}\right) + 4\, k\, L^2\, \sin^2 \: \alpha_1\, {\sin\!\left(\mathrm{\alpha_2}\right)}^4 + \\
    & \quad 14\, k\, L^2\, \sin^2 \: \alpha_1\, \sin^2 \: \alpha_2 + 12\, k\, L^2\, \sin^2 \: \alpha_1 - 4\, k\, L^2\, \sin\!\left(\mathrm{\alpha_1}\right)\, {\sin\!\left(\mathrm{\alpha_2}\right)}^3 - 8\, k\, L^2\, \sin\!\left(\mathrm{\alpha_1}\right)\, \sin\!\left(\mathrm{\alpha_2}\right) + \\
    & \quad 6\, k\, L^2\, {\sin\!\left(\mathrm{\alpha_2}\right)}^4 + 12\, k\, L^2\, \sin^2 \: \alpha_2 + 6\, k\, L^2
\end{align*}
\end{small}

\bibliographystyle{IEEEtran}
\bibliography{SK_RNB_KSP_DC_CDC2018}
\end{document}